\documentclass[epj]{webofc}
\usepackage{psfrag,pstool}
\usepackage[varg]{txfonts}

\newcommand{\order}{{\cal O}}

\definecolor{darkcyan}{rgb}{0.0, 0.55, 0.55}
\definecolor{darkred}{rgb}{0.55, 0.0, 0.0}
\definecolor{darkmagenta}{rgb}{0.55, 0.0, 0.55}

\wocname{EPJ Web of Conferences}
\woctitle{CONF12}

\begin{document}

\selectlanguage{english}
\title{Review of experimental and theoretical status of the proton radius puzzle}

\author{Richard J. Hill \inst{1,2,3,4}\fnsep\thanks{\email{richardhill@perimeterinstitute.ca}}
}

\institute{ Perimeter Institute for Theoretical Physics, Waterloo, ON, N2L 2Y5 Canada 
  \and
  Fermi National Accelerator Laboratory, Batavia, Illinois 60510, USA
  \and
  TRIUMF, 4004 Wesbrook Mall, Vancouver, BC, V6T 2A3 Canada
}

\abstract{ The discrepancy between the measured Lamb shift in muonic
  hydrogen and expectations from electron-proton scattering and
  regular hydrogen spectroscopy has become known as the proton radius
  puzzle, whose most "mundane" resolution requires a $> 5
  \sigma$ shift in the value of the fundamental Rydberg constant.  I
  briefly review the status of spectroscopic and scattering
  measurements, recent theoretical developments, and implications for
  fundamental physics.  }

\maketitle

\section{Introduction \label{sec:intro}}

The so-called proton radius puzzle is the $5.6 \sigma$ discrepancy
between the proton electric charge radius $r_E^p = 0.8751(61)$
measured from a combination of electron scattering and (regular,
electronic) hydrogen spectroscopy~\cite{Mohr:2015ccw}, and the radius
$r_E^p=0.84087(26)(29)$ measured from muonic hydrogen
spectroscopy~\cite{Antognini:1900ns}.%
\footnote{ These numbers correspond to the CODATA 2014 adjustment of
  constants, and the updated 2013 CREMA analysis.  The CODATA 2010
  value from electron measurements~\cite{Mohr:2012tt}, $r_E^p = 0.8775(51)$, and the
  original 2010 CREMA analysis~\cite{Pohl:2010zza}, $r_E^p =
  0.84184(36)(56)$, yielded a discrepancy of $6.9\sigma$.}
\footnote{For earlier reviews, see Refs.~\cite{Pohl:2013yb,Carlson:2015jba}.}

The large size of this discrepancy and its surprising appearance in
seemingly well-known systems, have motivated numerous theoretical and
experimental efforts  across particle, nuclear and atomic physics.
This talk begins by outlining the experimental basis for the puzzle in
Sec.~\ref{sec:outline}.  Section~\ref{sec:theory} then describes
theoretical issues that have received scrutiny since the emergence of
the puzzle, and Sec.~\ref{sec:expt} describes some emerging
experimental clues.  Section~\ref{sec:broader} describes a few of the
broader implications of the puzzle, and of work that has been
motivated by the puzzle.  Section~\ref{sec:outlook} provides an
outlook. 

\section{Outline of the puzzle \label{sec:outline}}

\begin{figure}[h]
  \hspace{-15mm}
  \psfragfig*[mode=nonstop,height=0.95\textwidth]{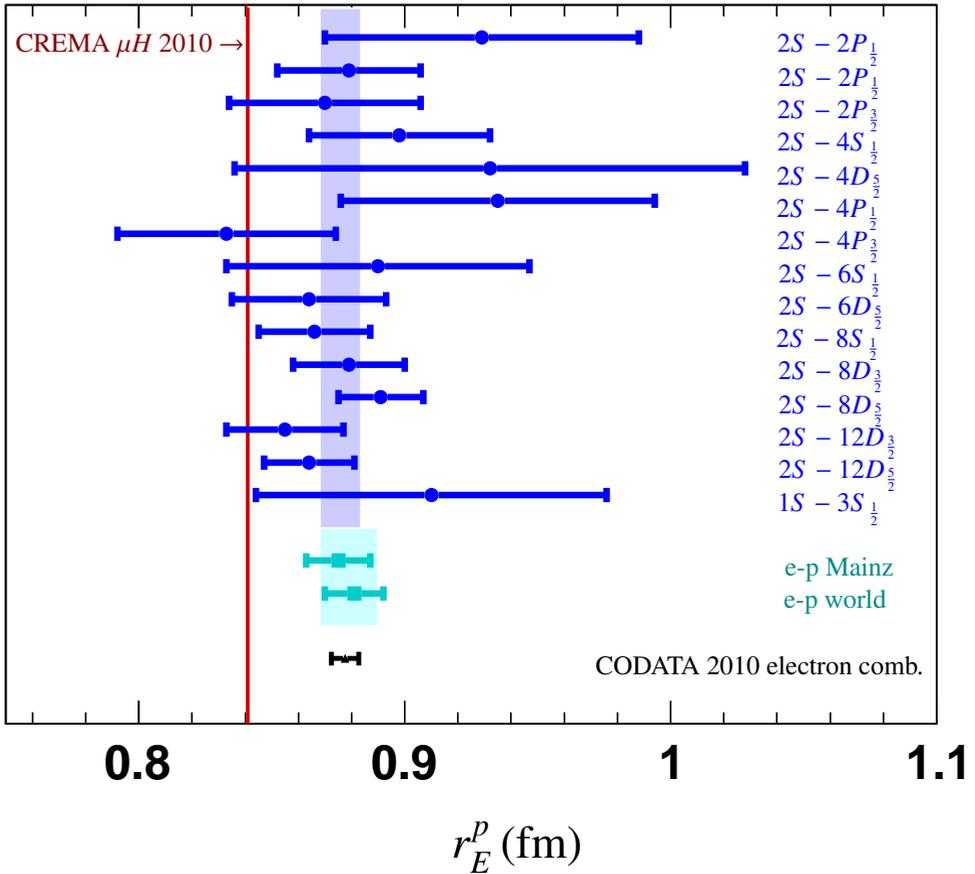}{
    \psfrag{a}{\color{blue}{$2S-2P_{\frac12}$} }
    \psfrag{b}{\color{blue}{$2S-2P_{\frac12}$} }
    \psfrag{c}{\color{blue}{$2S-2P_{\frac32}$} }
    \psfrag{d}{\color{blue}{$2S-4S_{\frac12}$} }
    \psfrag{e}{\color{blue}{$2S-4D_{\frac52}$} }
    \psfrag{f}{\color{blue}{$2S-4P_{\frac12}$} }
    \psfrag{g}{\color{blue}{$2S-4P_{\frac32}$} }
    \psfrag{h}{\color{blue}{$2S-6S_{\frac12}$} }
    \psfrag{i}{\color{blue}{$2S-6D_{\frac52}$} }
    \psfrag{j}{\color{blue}{$2S-8S_{\frac12}$} }
    \psfrag{k}{\color{blue}{$2S-8D_{\frac32}$} }
    \psfrag{l}{\color{blue}{$2S-8D_{\frac52}$} }
    \psfrag{m}{\color{blue}{$2S-12D_{\frac32}$} }
    \psfrag{n}{\color{blue}{$2S-12D_{\frac52}$} }
    \psfrag{o}{\color{blue}{$1S-3S_{\frac12}$} }
    \psfrag{p}{\color{darkcyan}{\hspace{0mm} e-p Mainz
    }}
    \psfrag{q}{\color{darkcyan}{\hspace{0mm} e-p world
    }}
    \psfrag{r}{\color{black}{\hspace{-25mm} CODATA 2010 electron comb.
    }}
    \psfrag{x}{\LARGE \color{black}{$r_E^p$\,(\rm fm)}} 
    \psfrag{xx}{ \color{darkred}{\hspace{-3mm}CREMA $\mu H$ 2010 $\rightarrow$} }
    }
  \caption{
    \label{fig:2010}
    Status of the proton radius puzzle circa 2010.  Blue data points
    denote various hydrogen intervals that are combined with the
    $1S-2S$ interval to solve for $r_E^p$ (datapoints as in
    Ref.~\cite{Beyer:2013jla}); blue band is the hydrogen average
    from Ref.~\cite{Mohr:2012tt}.  Cyan data points are electron-proton
    scattering determinations circa 2010 from Mainz A1 collaboration
    data~\cite{Bernauer:2010wm} and from other world
    data~\cite{Zhan:2011ji}; cyan band is the electron-proton
    scattering average from Ref.~\cite{Arrington:2015ria}. 
    The black data point represents the 2010
    CODATA~\cite{Mohr:2012tt} combination of hydrogen and
    electron-proton scattering determinations.
    The vertical red band is the 2010 CREMA determination from muonic
    hydrogen~\cite{Pohl:2010zza}.
  }
\end{figure}

The hydrogen spectrum depends on the Rydberg constant $R_\infty$ and
the proton charge radius $r_E^p$, schematically as~\cite{2016arXiv160703165P}
\begin{align}\label{eq:En}
E_{n,\ell} \sim -{R_\infty \over n^2} + \delta_{\ell 0} {\left(r_E^p\right)^2 \over n^3} \,, 
\end{align}
where $E_{n,\ell}$ is the energy for state of principle and angular
quantum numbers $n,\ell$.  The uncertainty on $r_E^p$, as presently
determined by electron-proton scattering and hydrogen spectroscopy,
limits the precision for $R_\infty$ that can be obtained using
Eq.~(\ref{eq:En}) and the precisely measured 1S-2S hydrogen
interval~\cite{Parthey:2011lfa,Matveev:2013orb}.  One motivation to
measure the Lamb shift (i.e., the 2P-2S interval) in muonic hydrogen
is that it can provide a precise determination of $r_E^p$, and hence a
more precise determination of $R_\infty$.   The surprising result in
2010 from the CREMA collaboration was a determination of the Lamb
shift~\cite{Pohl:2010zza} with an inferred charge radius not only much
more precise than, but in sharp tension with, results from both
regular hydrogen spectroscopy and electron-proton scattering.
Correspondingly, the Rydberg constant determined using the charge
radius inferred from the Lamb shift was in sharp tension with previous
determinations.  The measurements representing the proton radius
puzzle, circa 2010, are displayed in Fig.~\ref{fig:2010}.

\section{Status of some theory issues \label{sec:theory}}

I summarize here some recent progress regarding theoretical issues
impacting the proton radius puzzle.  I focus on the issues of form
factor nonlinearities and radiative corrections in electron-proton
scattering; and in higher-order proton structure effects in muonic
hydrogen.

\subsection{Electron-proton scattering: theory issues}

The proton charge radius is defined by the slope of the electric
charge form factor of the proton: 
\begin{align}\label{eq:raddef} 
  \frac16 \left(r_E^p \right)^2 \equiv {d\over dq^2} \bigg|_{q^2=0} G_E^p(q^2) \,, 
\end{align}
where $q^2=-Q^2$ is the invariant momentum transfer of the scattering
process.  Two important issues surrounding the determination of the
charge radius from scattering data are the treatment of form factor
nonlinearities, and the treatment of radiative corrections. 

\subsubsection{Form factor nonlinearities}

\begin{figure}[htb]
  \centering
  \includegraphics[height=0.6\textwidth]{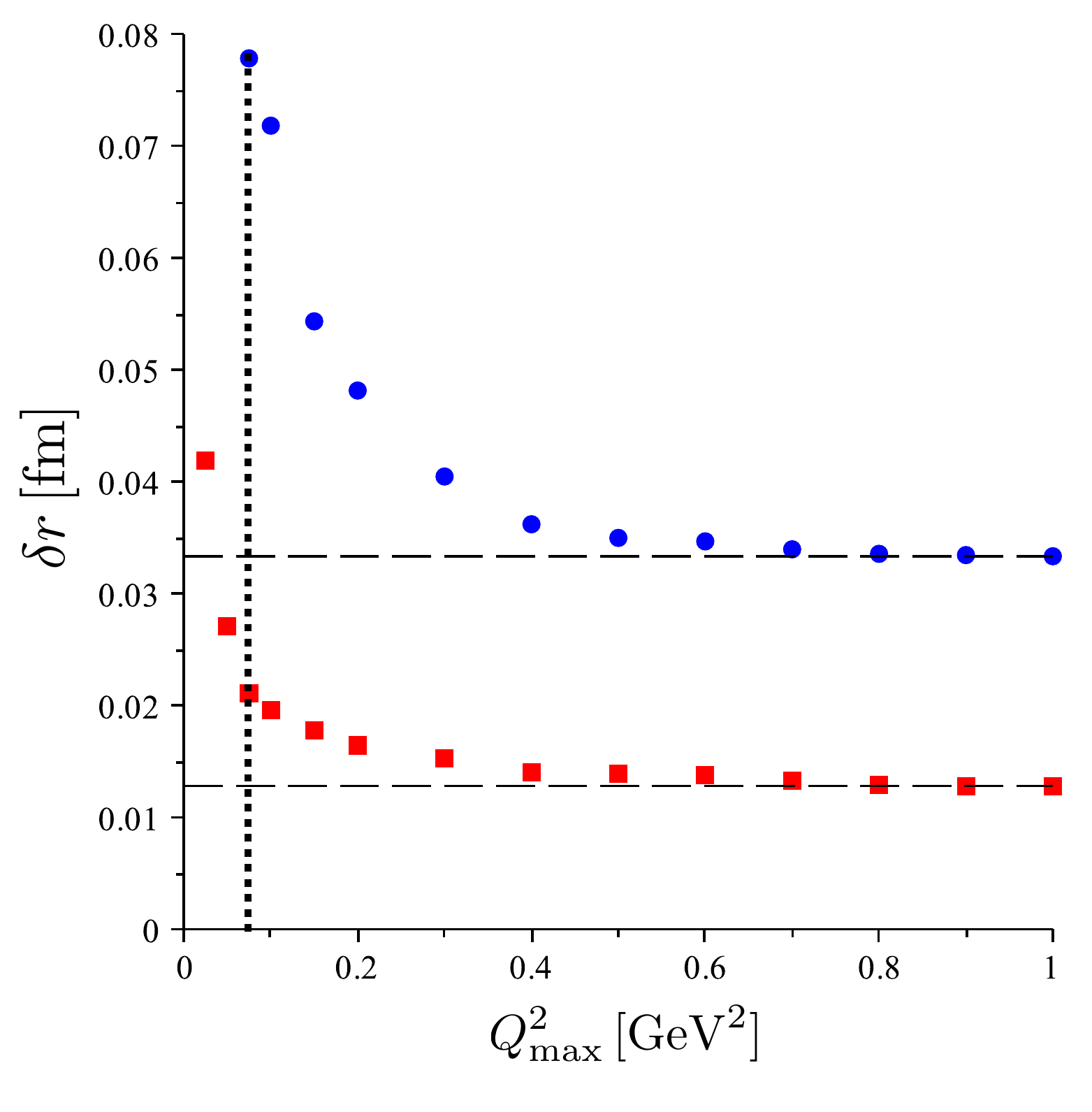}
  \vspace{-5mm}
  \caption{\label{fig:sens} Statistical error on $r_E^p$ as a function
    of the maximum momentum transfer retained in the fit, $Q^2_{\rm
      max}$,  for the 1422 point A1 MAMI dataset (red squares) and for
    the complementary world cross section and polarization dataset
    (blue circles).  The horizontal dashed lines are large-$Q^2_{\rm
      max}$ asymptotes.  The vertical dotted line  represents the
    limit $Q^2_{\rm max}=4 m_\pi^2$ beyond which the two-pion threshold
    introduces nonanalytic structure.  For details, see
    Ref.~\cite{Lee:2015jqa}.  } 
\end{figure} 
  
Figure~\ref{fig:sens} illustrates an essential feature of fits to
electron-proton scattering data.
Since the radius is defined by the
slope of the form factor, Eq.~(\ref{eq:raddef}), a finite range of
momentum transfer must be included.  Care must then be taken to account for
nonlinearities in form factor shape.  In fact, the situation is more
dire than simply needing to account for higher-order terms in the
$Q^2$ Taylor expansion of $G_E^p$: in order to obtain relevant precision
with current data (i.e., with an error small compared to the $\sim
0.034\,{\rm fm}$ radius anomaly) the required $Q^2_{\rm max}$ is larger than the radius of convergence for the form
factor!

\begin{figure}[htb]
  \centering
    \hspace{-15mm}
    \psfragfig*[mode=nonstop,height=0.35\textwidth]{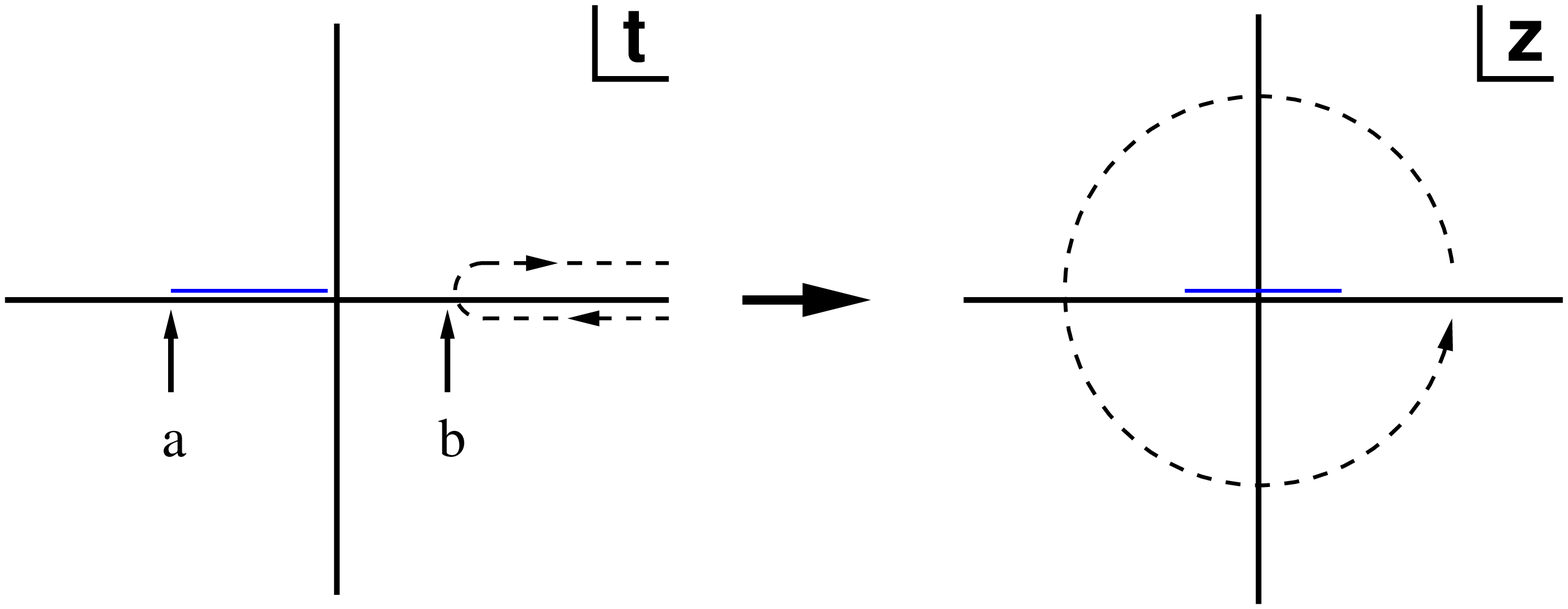}{
    \psfrag{t}{$q^2$}
    \psfrag{z}{$z$}
    \psfrag{a}{$-Q^2_{\rm max}$}
    \psfrag{b}{$t_{\rm cut}$}
    }
    \vspace{-5mm}
  \caption{\label{fig:z}
    Variable transformation $q^2 \to z(q^2)$ mapping the cut plane
    to the unit circle.  The physical region for spacelike scattering, denoted in blue, is
    $-Q^2_{\rm max} < q^2 < 0$.  The singularity corresponding to pion production
    in the crossed channel is $t_{\rm cut}=4 m_\pi^2$.  
  }
\end{figure}

Happily, this problem is readily solved: as illustrated in
Fig.~\ref{fig:z}, a variable change mapping the cut plane onto the
unit circle ensures convergence of a Taylor expansion in the new
variable throughout the entire domain of analyticity:%
\footnote{Formalism for $z$ expansion and nucleon form factors is
  described in  Refs.~\cite{Hill:2010yb,Bhattacharya:2011ah}, and
  several applications are  found in
  Refs.~\cite{Lorenz:2014vha,Epstein:2014zua,Lee:2015jqa,Bhattacharya:2015mpa}.
  Related formalism and applications may be found
  in~\cite{Hill:2006ub, Bourrely:1980gp,
    Boyd:1994tt,Boyd:1995sq,Lellouch:1995yv,Caprini:1997mu,Arnesen:2005ez,
    Becher:2005bg,Hill:2006bq,Bourrely:2008za,Bharucha:2010im,Amhis:2014hma,Bouchard:2013pna,Bailey:2015dka,Horgan:2013hoa,Lattice:2015tia,Detmold:2015aaa}.
}
\begin{align}\label{eq:zexp}
  G_E^p(q^2) = \sum_{k=0}^{\infty} a_k [z(q^2)]^k \,, \qquad z(q^2) =
  \frac{\sqrt{t_{\rm cut} - q^2} - \sqrt{t_{\rm cut} -
      t_0}}{\sqrt{t_{\rm cut} - q^2} + \sqrt{t_{\rm cut} - t_0} } \,. 
\end{align}
Moreover, the expansion is systematically improvable, and the error
due to truncating the expansion at a given order is reliably
estimated.%
\footnote{ The maximum size of $|z(q^2)|$ is bounded in a given
  kinematic window, and the coefficients $a_k$ are dimensionless order
  unity numbers.  }

A range of parameterizations has been used to describe the elastic vector
form factors of the proton.%
\footnote{
  For a discussion and references, see Ref.~\cite{Hill:2010yb}.  
}
In many cases, these parameterizations are in conflict with known
properties of the form factors inherited from QCD.  For example,
as discussed around Fig.~\ref{fig:sens},
a Taylor expansion around $q^2=0$ is valid only
up to $q^2=4 m_\pi^2$ where pion production in the crossed channel
corresponds to a branch point singularity.   As another example,
a Pad\'{e} approximation of continued fractions can be justified
only when the spectral function appearing in the dispersive
representation of the form factor, ${\rm Im}\, G_E(q^2)$, is positive
definite.%
\footnote{
  A positive spectral function would predict an asymptotic
  scaling at large momentum transfer $\sim 1/Q^2$, in conflict
  with the known $\sim 1/Q^4$ behavior.
  For an application where this positivity condition {\it is}
  satisfied, see Ref.~\cite{Chakraborty:2014mwa}.}
Such parameterizations, with sufficiently many parameters, may be able
to fit the available data.  However, without a priori control over the
number of relevant parameters, there is an inevitable arbitrariness in
deciding how many parameters to keep, and a complicated analysis is
required to understand the interplay with statistical and systematic
experimental errors. 

\begin{figure}[htb]
  \centering
    \hspace{-15mm}
  \psfragfig*[mode=nonstop,height=0.75\textwidth]{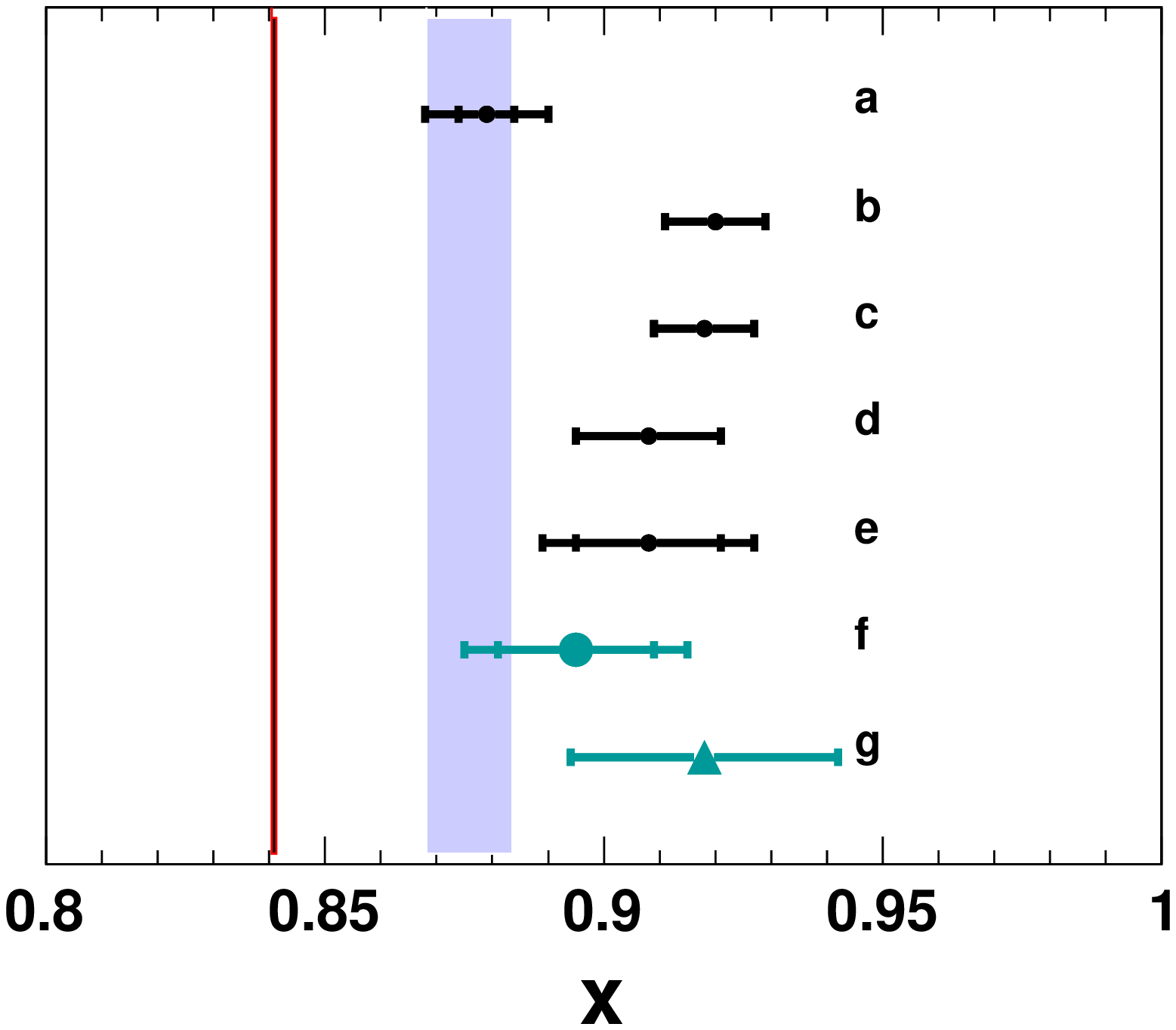}{
    \psfrag{a}{A1 spline}
    \psfrag{b}{Bounded $z$ exp.}
    \psfrag{c}{+ Hadronic TPE}
    \psfrag{d}{\hspace{-10mm} Rebin, 0.3\%-0.4\% syst.}
    \psfrag{e}{+0.4\% corr. syst.}
    \psfrag{f}{$Q^2_{\rm max} = 0.5\,{\rm GeV}^2$}
    \psfrag{g}{other world data}
    \psfrag{x}{\LARGE \color{black}{$r_E^p$\,(\rm fm)}}
  }
  \vspace{-5mm}
  \caption{
    \label{fig:reanalysis}
    Radius extraction from 2010 Mainz A1 collaboration data.  The topmost
    point is the original A1 analysis for reference.  The next five points
    represent $z$ expansion fits under a series of modifications as
    described in the text, culminating in the final result given by
    the large cyan circle.  The final point (large triangle) represents
    the same analysis
    applied to other world data.  From Table~XIV of Ref.~\cite{Lee:2015jqa}.
    The vertical blue and red bands are the regular hydrogen and
    muonic hydrogen results reproduced from Fig.~\ref{fig:2010}.
  }
\end{figure}

With theoretical control over form factor nonlinearities, we may
revisit the extraction of the charge radius from scattering data.
Figure~\ref{fig:reanalysis} shows several steps in a reanalysis of the
1422 point Mainz A1 dataset (see Ref.~\cite{Lee:2015jqa}, Table~XIV).
The topmost point (``A1 spline'')
displays the original A1 analysis result,%
\footnote{ As discussed in Ref.~\cite{Lee:2015jqa},  this results from
  adding different systematic errors  linearly~\cite{Priv_Distler},
  compared to the quadrature sum advocated in
  Ref.~\cite{Bernauer:2013tpr}.  }
employing the entire $Q^2$ range (up to $1\,{\rm GeV}^2$) and a cubic
spline fit~\cite{Bernauer:2013tpr}. The next point (``Bounded $z$
exp.'') displays the corresponding result using precisely the same
data and errors as the first point, but replacing the spline
parameterization by the $z$ expansion [with standard statistical
priors on the coefficients $a_k$ in Eq.~(\ref{eq:zexp})].  The next four
points show the impact of using a more conventional radiative
correction model (``+ Hadronic TPE''); rebinning data taking at
identical kinematics (``Rebin, 0.3\%-0.4\% syst.'');%
\footnote{ The rescaling analysis by which kinematically uncorrelated
  systematic errors were estimated in Ref.~\cite{Bernauer:2013tpr}
  would otherwise unintentionally drive these errors to zero with
  repeated measurements at the same kinematics.  }
including a larger error budget to account for known sources of
correlated systematics (``+0.4\% corr. syst.''); and choosing $Q^2_{\rm max}=0.5\,{\rm
  GeV}^2$ to maximize radius sensitivity but minimize potential
large-$Q^2$ systematics (cf. Fig.~\ref{fig:sens}).  The final point in the figure displays the
result of a similar analysis applied to other world data (not
including the Mainz A1 dataset).   It is readily seen that form factor
shape assumptions can have dramatic consequences for radius
extractions.

\begin{figure}[htb]
  \centering
  \includegraphics[height=0.35\textwidth]{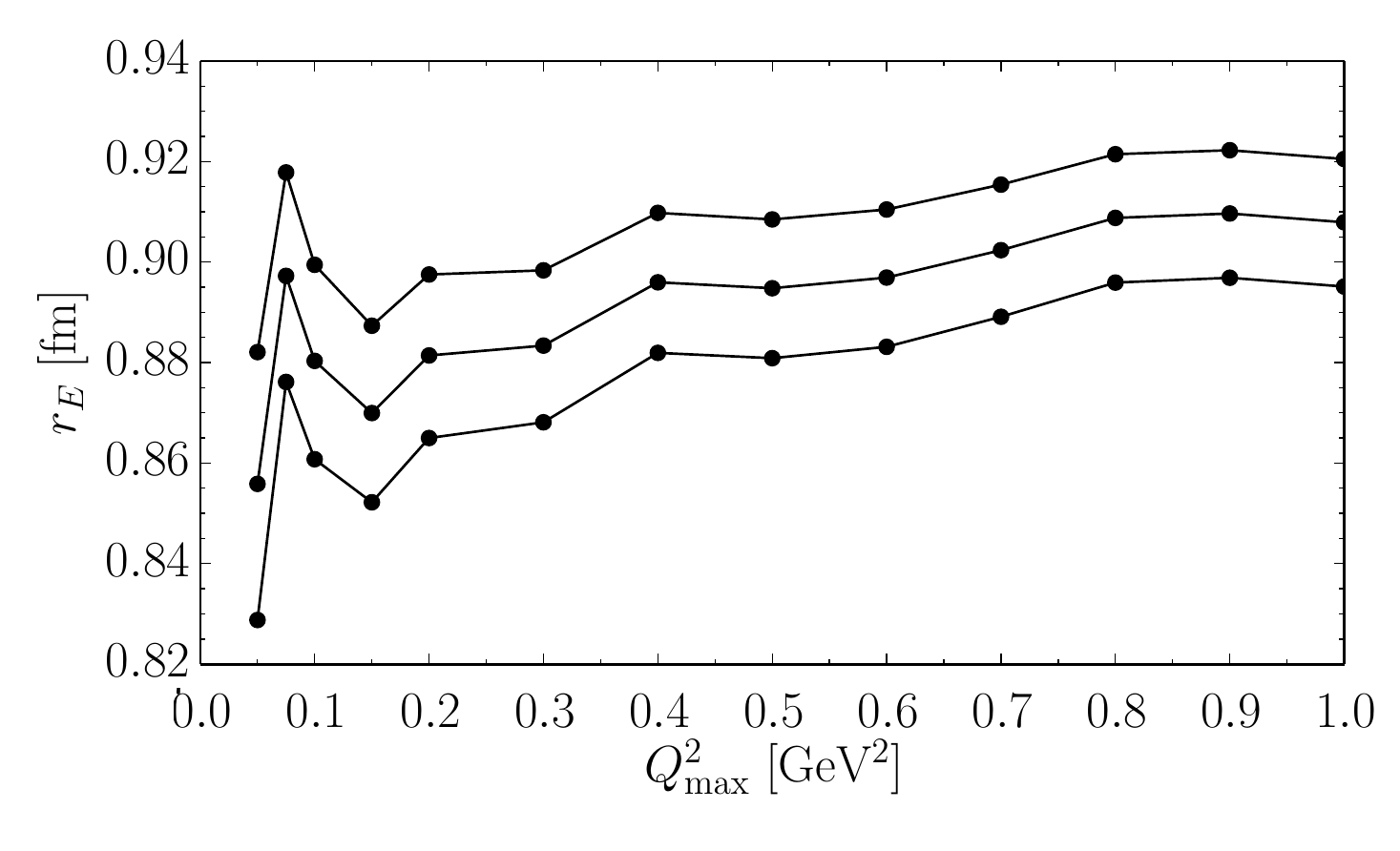}
  \includegraphics[height=0.35\textwidth]{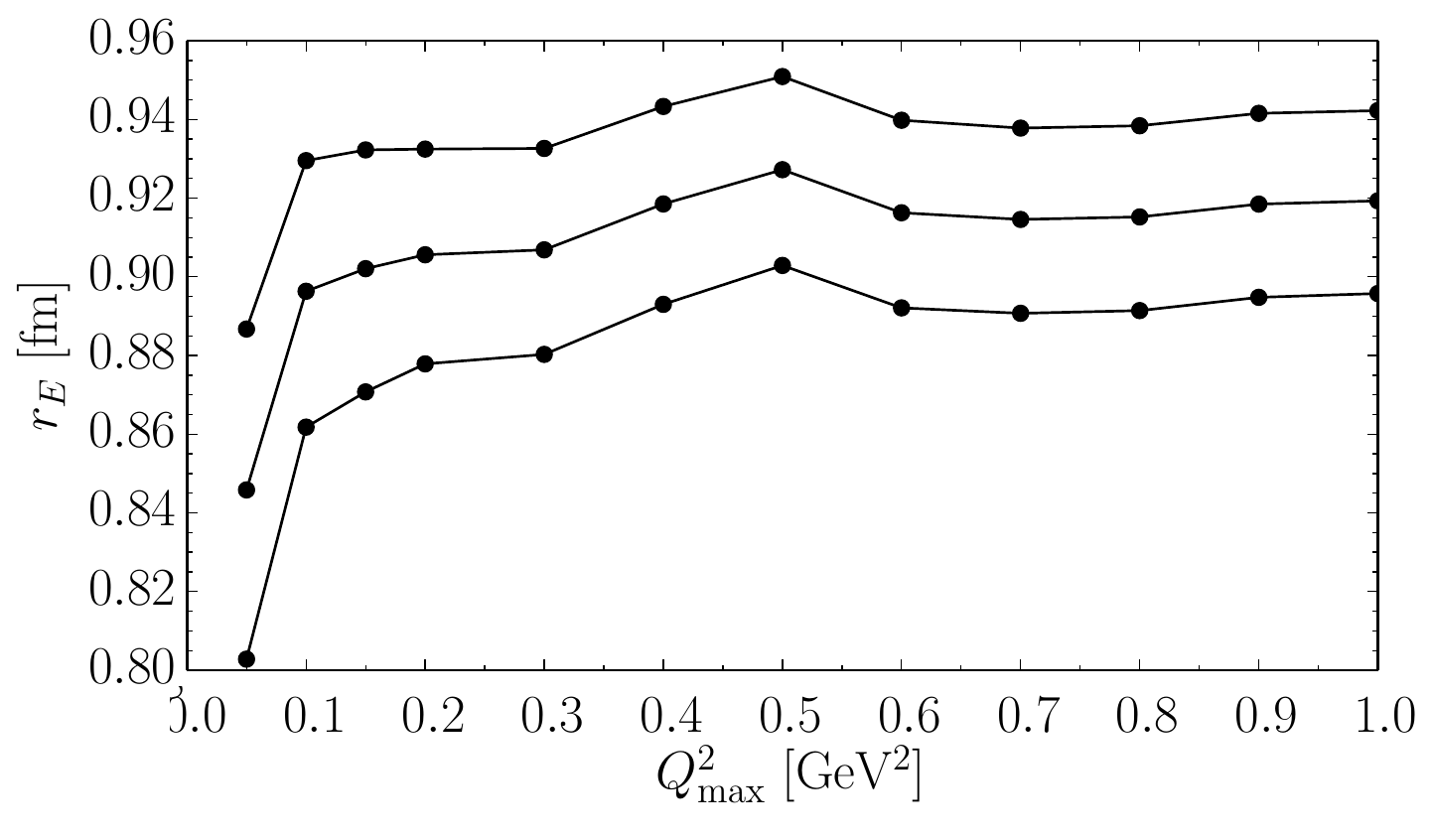}
  \caption{\label{fig:Q2max} Proton electric charge radius
    as a function of maximum $Q^2$ retained in the fit to scattering data.
    The three curves represent central values and $1\sigma$ error bars. 
    Top: 2010 Mainz A1 dataset (statistical and uncorrelated systematic errors only).
    Bottom: other world data.  From Ref.~\cite{Lee:2015jqa}. 
  }
\end{figure} 

The decomposition of the high statistics Mainz A1 collaboration
dataset~\cite{Bernauer:2013tpr} into subsets with independent floating
normalization parameters has been used to minimize the impact of
poorly-measured systematic effects that primarily impact overall cross
section normalization.  A variant of this analysis in
Ref.~\cite{Lee:2015jqa} has been used to quantify how large a missing
or mismodeled systematic effect would need to be in order to impact
the radius extraction. In short, there would need to be: a variation
larger than $\sim 0.4\%$ over data subsets;%
\footnote{ The data is divided into 18 subsets corresponding to
  combinations of beam energy and spectrometer configuration.  }
or a variation of a more extreme functional form than those
considered; or unaccounted correlations between different subset
variations.

Figure~\ref{fig:Q2max} displays the dependence of the extracted radius
on the range of $Q^2$ considered.   There is mild tension between fits
to the entire $Q^2$ range below $1\,{\rm GeV}^2$, and fits restricted
to low-$Q^2$.   While not of high statistical significance, the
appearance of a similar feature in fits to independent datasets may be
suggestive a common theoretical systematic.  It does not escape
attention that the low-$Q^2_{\rm max}$ fits have central $r_E^p$ value
closer to the $r_E^p$ from muonic hydrogen (albeit with large error).
Moreover, radiative corrections are known to be enhanced at large
$Q^2$, owing to large logarithms, $\sim \log(Q^2/m_e^2)$ in the
perturbative expansion.  Let us revisit the status of these
corrections. 

\subsubsection{Radiative corrections}

\begin{figure}[htb]
  \centering
  \includegraphics[height=0.35\textwidth]{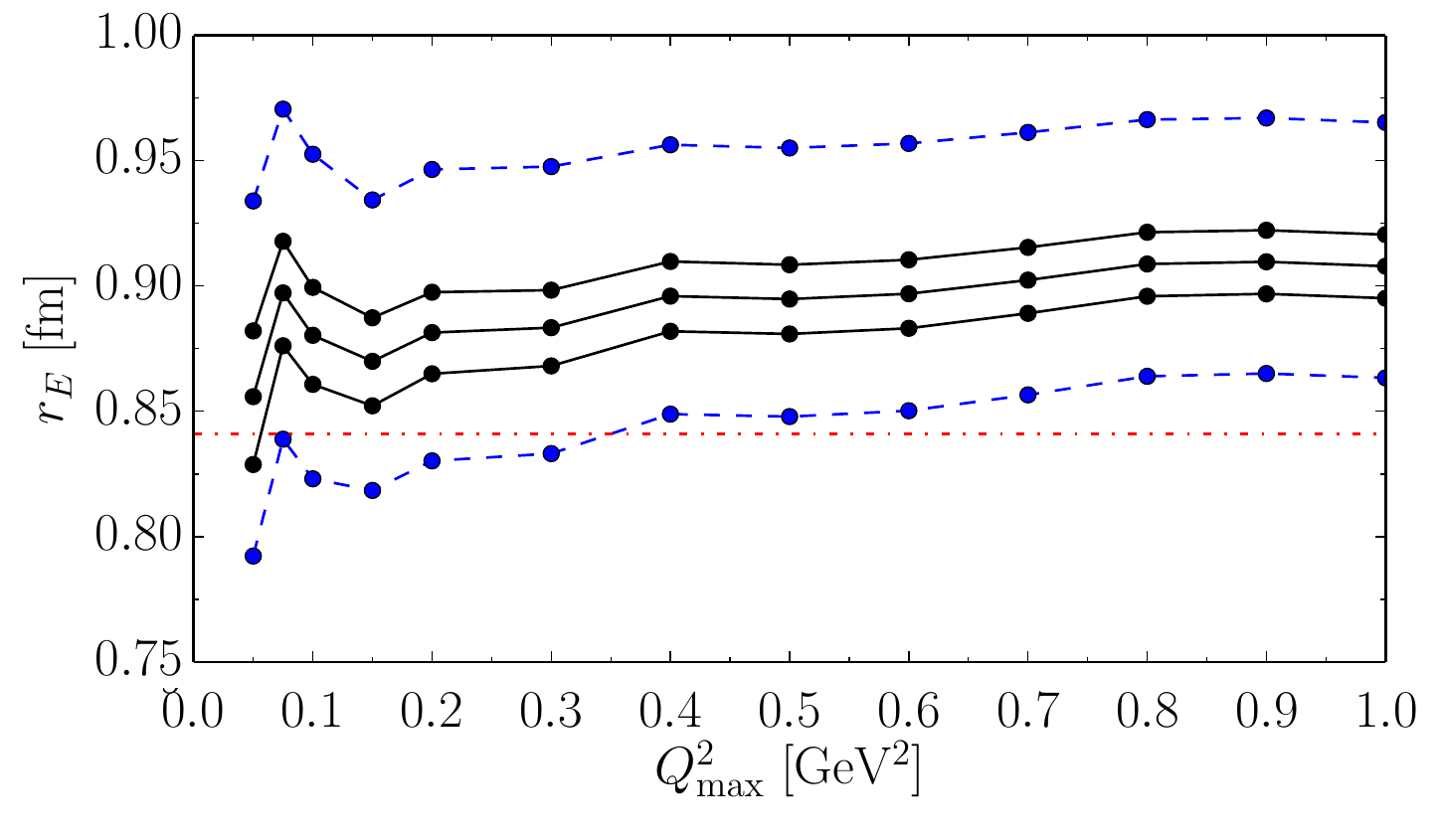}
  \includegraphics[height=0.35\textwidth]{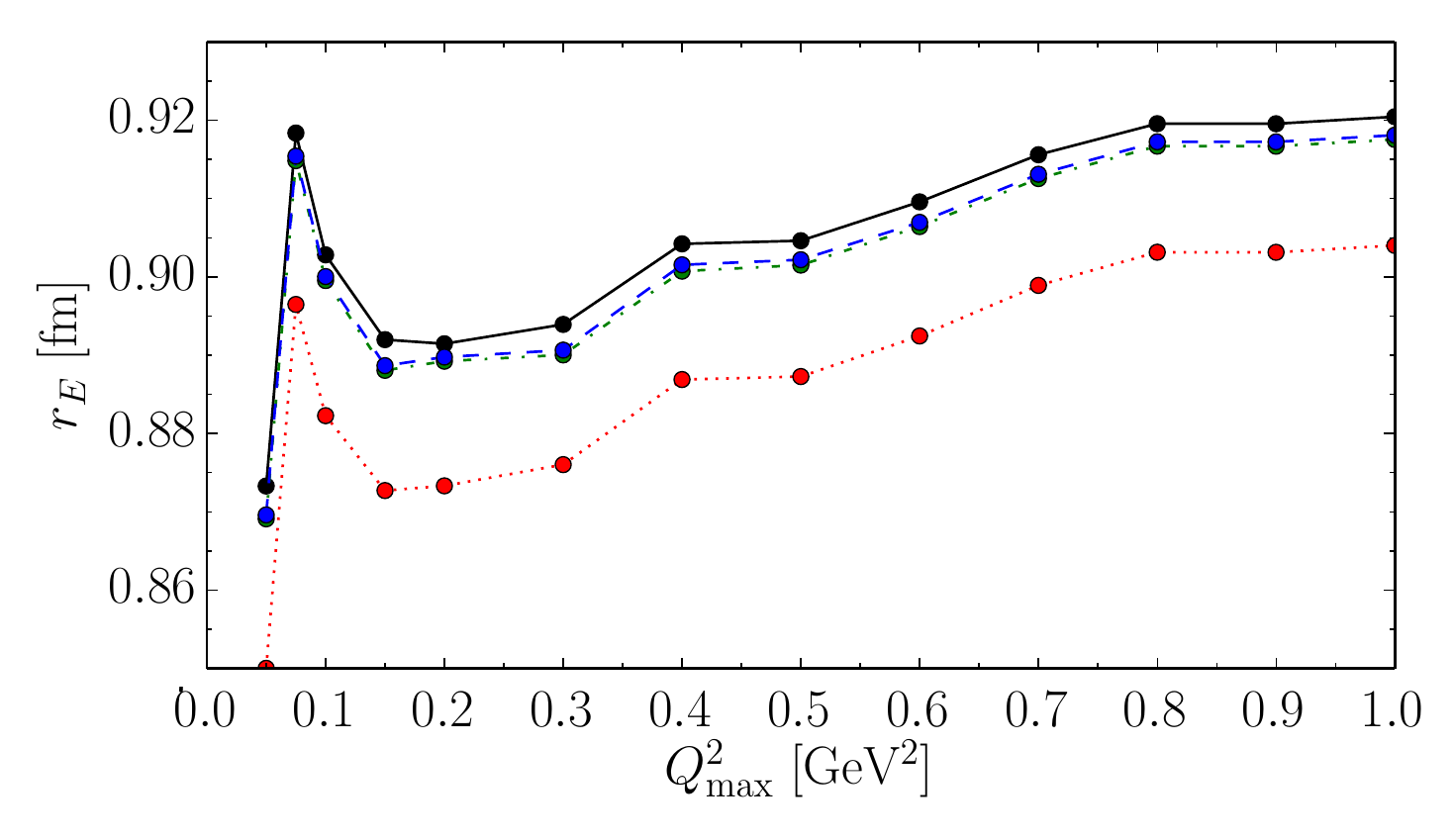}
  \caption{\label{fig:impact}Potential impact of radiative
    corrections.  Top: black solid lines are from the top plot of
    Fig.~\ref{fig:Q2max}.  The blue dashed lines represent an error
    band from factorization scale variation in the radiative corrections,
    where only leading logarithms
    are controlled.  The red dotted line is the central value
    from muonic hydrogen.  Bottom: the bottom, red, dotted line is the
    radius extracted with vanishing hard TPE correction.  The
    remaining lines are results using different models for hard TPE.
    From Ref.~\cite{Lee:2015jqa}. 
  }
\end{figure} 

Here I review the definition of the charge radius in the presence of QED 
radiative corrections, and discuss the status of soft and hard radiative
corrections to the scattering process.  

The definition (\ref{eq:raddef}) assumes a choice for Born form
factor.%
\footnote{ For a discussion of definitions including QED radiative
  corrections, see Appendix~B of Ref.~\cite{Hill:2016gdf}.   }
In modern effective field theory language, $G_E$ is the hard
contribution in the factorization formula for the onshell form factor.
It should be noted that a number of different conventions exist in the
literature for defining the radius when accounting for radiative
corrections.  The convention advocated in Refs.~\cite{Hill:2011wy,
  Hill:2016gdf} is defined by the $\overline{\rm MS}$ renormalization
scheme for the soft function, and has the property that the radius for
a point particle vanishes in the presence of first order radiative
corrections. 

\begin{figure}[htb]
  \centering
  \hspace{-5mm}
  \includegraphics[height=0.43\textwidth]{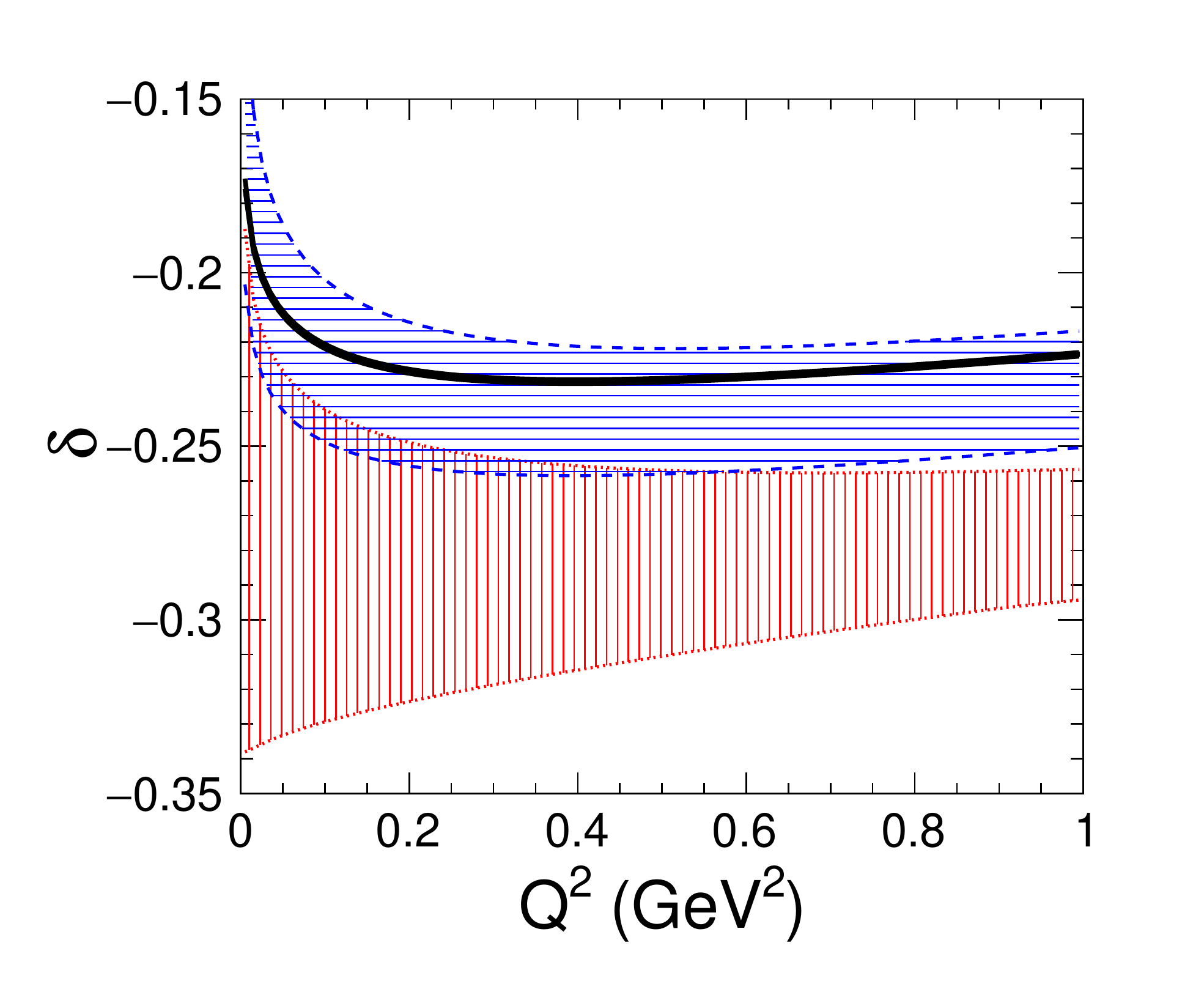}
  \hspace{-5mm}
  \includegraphics[height=0.43\textwidth]{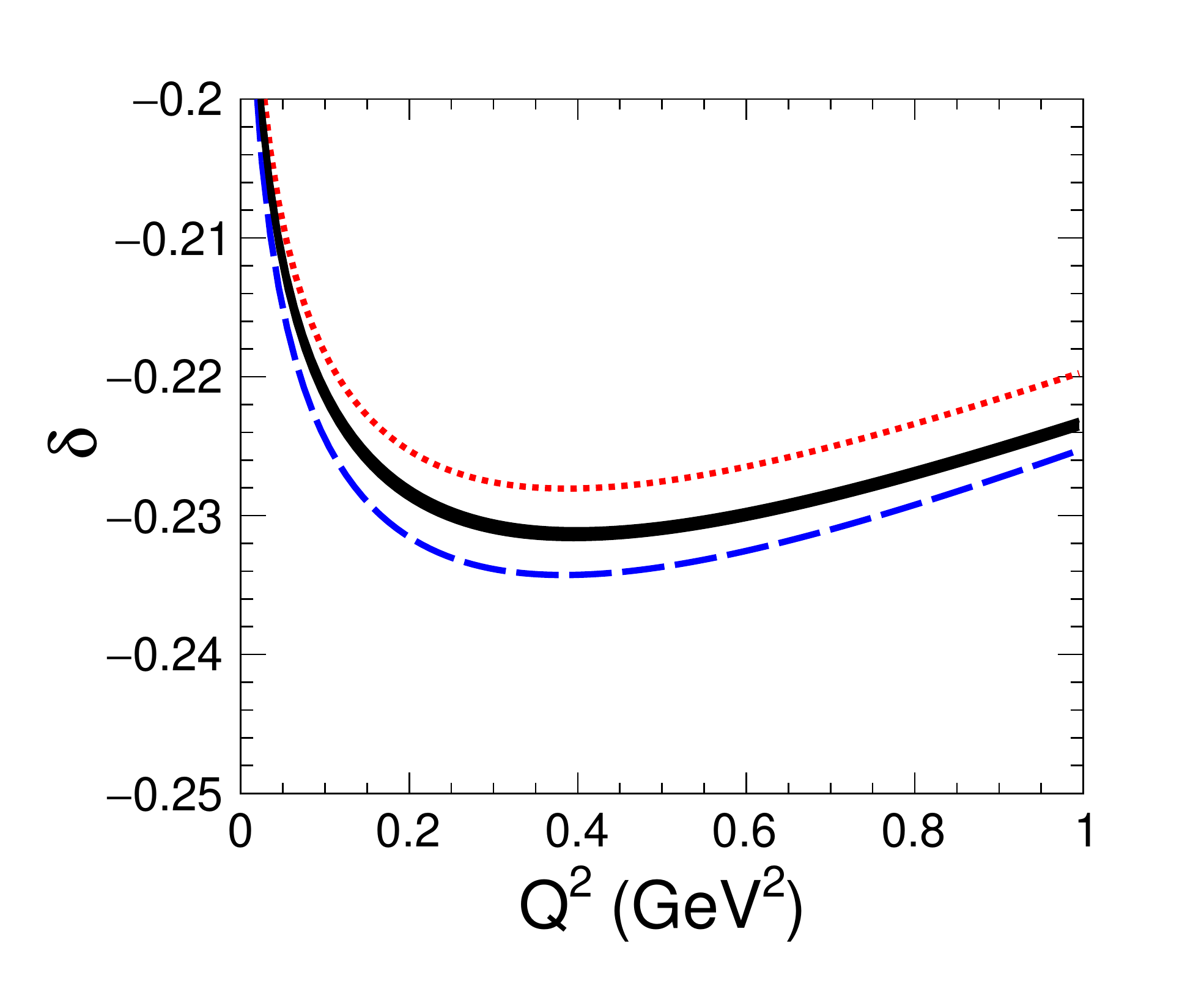}
  \vspace{-5mm}
  \caption{\label{fig:radcor} Radiative correction factor $\delta$
    in Eq.~\ref{eq:factor}, at electron energy $E=1\,{\rm GeV}$
    and bremsstrahlung cut $\Delta E = 5\,{\rm MeV}$.
    LHS: red vertical hashed, blue horizontal hashed, and black solid bands
    represent
    leading logarithm, next-to-leading logarithm and next-to-leading order
    calculations.  RHS: black is the same as LHS; red dotted and blue dashed
    give two versions of the previously-employed exponentiation ansatz.  From Ref.~\cite{Hill:2016gdf}.  
  }
  \vspace{-5mm}
\end{figure}

The complete cross section for the elastic scattering process 
including radiative corrections may be written,
\begin{align}\label{eq:factor}
  d\sigma \propto H \times J \times R \times S \equiv H(\mu=M) \times (1+\delta) \,,
\end{align}
where $H$ (hard contribution) is defined to contain the Born form factors and 
hard TPE, and the $J$ (jet), $R$ (remainder) and $S$ (soft) factors are
calculable in QED.  In the latter equality we have defined the scheme choice
for $H$ (at factorization scale $\mu=M$), and defined the radiative correction
factor $\delta$ appearing in Fig.~\ref{fig:radcor}. 

The product $JRS$ in Eq.~(\ref{eq:factor}) contains large logarithms,
$L \sim \log{Q^2/m_e^2}$.  For typical scattering kinematics, the
argument of the logarithm involves ratios of GeV to MeV scales.
Leading corrections are as large as $\sim 30\%$, and naively
subleading corrections must also be included.  In previous analyses,
an exponentiation ansatz has been
employed~\cite{Vanderhaeghen:2000ws,Bernauer:2013tpr} to account for
logarithmically enhanced terms at second- and higher-order in
perturtabive QED.   This procedure fails to capture subleading
logarithms, beginning at order $\alpha^2 L^3$.   The potential impact
of such subleading corrections is illustrated in
Fig.~\ref{fig:impact}, top, where an effective renormalization scale
variation is represented by the blue lines.  Clearly, corrections
beyond the leading terms must be controlled in order to exhibit a
discrepancy between the scattering data (black solid line) and muonic
hydrogen (red dotted line). 

The factor $H$ in Eq.~(\ref{eq:factor}) contains 
the hard two-photon exchange
contribution, which must be subtracted in order to isolate the Born
form factors of interest.  Figure~\ref{fig:impact}, bottom, displays
the radius extracted using 
several models for the hard contribution to
TPE~\cite{McKinley:1948zz,Blunden:2005ew,Blunden:2003sp,Carlson:2007sp,Arrington:2011dn,Tomalak:2016vbf}.  The variation
between red line (no hard correction) and other lines shows that the
total hard TPE contribution enters at a level comparable to the proton
radius anomaly.  Clearly this contribution must also be reliably
controlled. 

Figure~\ref{fig:radcor}, LHS, shows the complete calculation of the
radiative correction factor $\delta$~\cite{Hill:2016gdf} for an illustrative kinematic point,
displaying convergence of leading logarithms (red), next-to-leading logarithms (blue) and a
complete next-to-leading order calculation (black).
Figure~\ref{fig:radcor}, RHS, shows the same complete calculation
(black) compared to the previously-employed exponentiation ansatz
(red).  The difference between black and red curves is the impact of subleading
logarithms missing in the exponentiation ansatz. 

In previous work employing the exponentiation ansatz, 
there is an implicit choice of conflicting renormalization scales in the one-photon exchange and
two-photon exchange contributions that results in an ambiguity between
the red and blue lines in Fig.~\ref{fig:radcor}.%
\footnote{In particular, the usual convention for Born form factors corresponds
  to the scale choice $\mu=M$, $M$ the proton mass, while the usual (Maximon-Tjon)
  convention for hard TPE corresponds to the choice $\mu=Q$.} 
The difference between the red and blue curves is an (indirect) measure of hard
TPE model uncertainty.

These modifications to the radiative corrections impact cross sections at
the $\sim 0.5-1\%$ level over the kinematics of the A1 experiment.
While this is in tension with the assumed error budget
of $\sim 0.1-0.2\%$, a large part of the correction will be absorbed
by floating normalization parameters.  

\subsubsection{Interplay with other constraints}

A variety of alternative assumptions regarding fit functions and data
selections have been employed in the literature to extract the charge
radius from scattering data (for an incomplete survey, see 
Refs.~\cite{Sick:2012zz,Sick:2014sra,Griffioen:2015hta,Horbatsch:2015qda,Higinbotham:2015rja,
  Lorenz:2014vha,Lorenz:2014yda,Kraus:2014qua,Horbatsch:2016ilr,Sick:2017aor}).
Several analyses argue for a small radius obtained by retaining only
the low-$Q^2$ scattering data.  As Fig.~\ref{fig:sens} shows,
inclusion of only low-$Q^2$ data (e.g., where a Taylor expansion is
appropriate) cannot achieve sufficient precision to address the proton
radius puzzle without further information.%
\footnote{ Reference~\cite{Horbatsch:2016ilr} argues to supplement the
  low-$Q^2$ scattering data with constraints on curvature and
  higher-order derivatives of the form factor from chiral perturbation
  theory.   }
Beyond the question of statistical power, such an approach must assume
that any systematic impacting higher-$Q^2$ data does not invalidate
the lower-$Q^2$ data.  

Isospin decomposition of electron-proton and electron-neutron data can
be used to place somewhat tighter constraints on the form
factors~\cite{Belushkin:2006qa,Hill:2010yb}, since the isoscalar
threshold is larger: phrased in terms of $z$ expansion, the larger
threshold, $(3 m_\pi)^2$ versus $(2 m_\pi)^2$, implies a smaller
maximum size of $|z(q^2)|$, and faster convergence with fewer relevant
expansion coefficients~\cite{Hill:2010yb}.  Fits to the combined
proton and neutron data, together with $\pi \pi \to N\bar{N}$
constraints, have been argued to imply a smaller value of $r_E^p$.
However, this conclusion relies on modeling the spectral function at
inaccessible kinematics.%
\footnote{ See for example, Ref.~\cite{Hill:2010yb}, where inclusion
  of the model-independent part of the spectral function reduces
  uncertainty but does not dramatically alter the central value of the
  radius extracted from the considered data.  For recent analysis, see
  also Ref.~\cite{Hoferichter:2016duk}.   }
Beyond the question of model-dependence in the spectral functions, a
result consistent with muonic hydrogen again would demand that the
scattering data be effectively overruled by other constraints or
assumptions.  

\subsection{Muonic hydrogen: theory issues}

Muonic hydrogen is much more sensitive to proton structure than
ordinary hydrogen, due to the huge wavefunction enhancement,
\begin{align}
  { |\psi_{\mu H}(0)|^2 \over |\psi_{e H}(0)|^2 } \sim {m_\mu^3\over m_e^3} \sim (200)^3 \,. 
\end{align}
Relative to the leading $\sim m_\ell \alpha^2$ contribution, proton structure
impacts the muonic hydrogen spectrum $\sim (200)^2$ times more strongly than regular
hydrogen.  

\begin{figure}[htb]
  \centering
    \hspace{-15mm}
  \psfragfig*[mode=nonstop,height=0.25\textwidth]{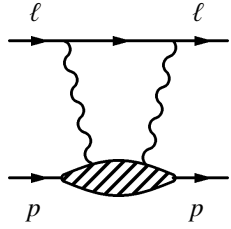}{
    \psfrag{e}{$\ell$}
    \psfrag{p}{$p$}
  }
  \vspace{-5mm}
  \caption{
    \label{fig:blob2}
    Two photon exchange contribution to forward scattering $\ell p \to \ell p$,
    where $\ell = e,\mu$. 
  }
  \vspace{-5mm}
\end{figure}

In addition to the radius, other proton structure effects enter, in particular
two-photon exchange contributions.
The matching condition from a relativistic QCD theory of quarks and gluons 
onto the low energy Hamiltonian is represented by the box diagram of
Fig.~\ref{fig:blob2}, which is given by an integral 
over the forward Compton amplitude of the proton.  
The relevant proton structure is contained in the spin-averaged
Compton amplitude, described by two invariant amplitudes,
$W_i(\nu,Q^2)$, $i=1,2$.  These amplitudes are determined by
elastic and inelastic scattering data using dispersion relations. 
However one of the dispersion relations does not converge, and requires
a subtraction,
\begin{eqnarray}\label{dispersion}
  W_1(\nu,Q^2) &=& W_1(0,Q^2)+
  {\nu^2\over \pi} \int_{0}^\infty { d\nu^{\prime 2}} 
\,{ {\rm Im}W_1(\nu^\prime, Q^2) \over \nu^{\prime 2} (\nu^{\prime 2} - \nu^2) }  \,,
\nonumber\\\nonumber\\
W_2(\nu,Q^2) &=&
{1\over \pi} \int_{0}^\infty { d\nu^{\prime 2} }
\,{ {\rm Im}W_2(\nu^\prime, Q^2) \over \nu^{\prime 2} - \nu^2}
\,.
\end{eqnarray} 
The subtraction term, $W_1(0,Q^2)$ is a major source of uncertainty,
since it is not directly determined by scattering data.

\begin{figure}[h]
  \centering
  \includegraphics[height=0.5\textwidth]{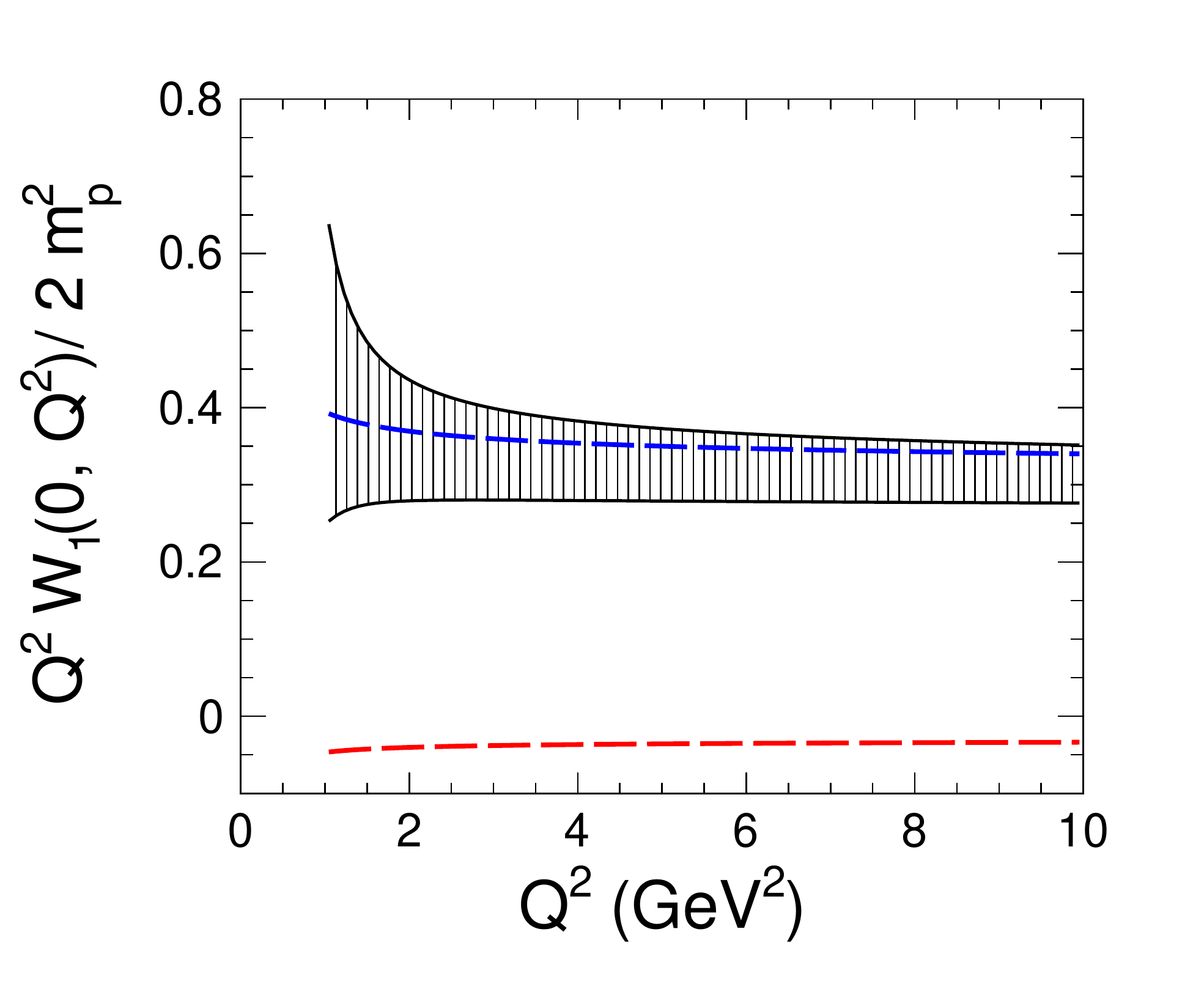}
  \vspace{-5mm}
  \caption{Leading OPE prediction for $W_1(0,Q^2)$.  The bottom red dashed line, and top blue dashed
    line represent central values of the spin-0 and spin-2 contributions.   The hatched band represents
    the total including perturbative and hadronic uncertainty.  From Ref.~\cite{Hill:2016bjv}.
  }
\end{figure} 

The subtraction function can be computed in various regimes.
At low $Q^2 \ll m_\pi^2$ the Taylor expansion of $W_1(0,Q^2)$ is determined by
Wilson coefficients of NRQED~\cite{Hill:2011wy}; these coefficients are in turn determined
by data or by nonperturbative QCD calculations.

\begin{figure}[htb]
  \centering
  \includegraphics[height=0.5\textwidth]{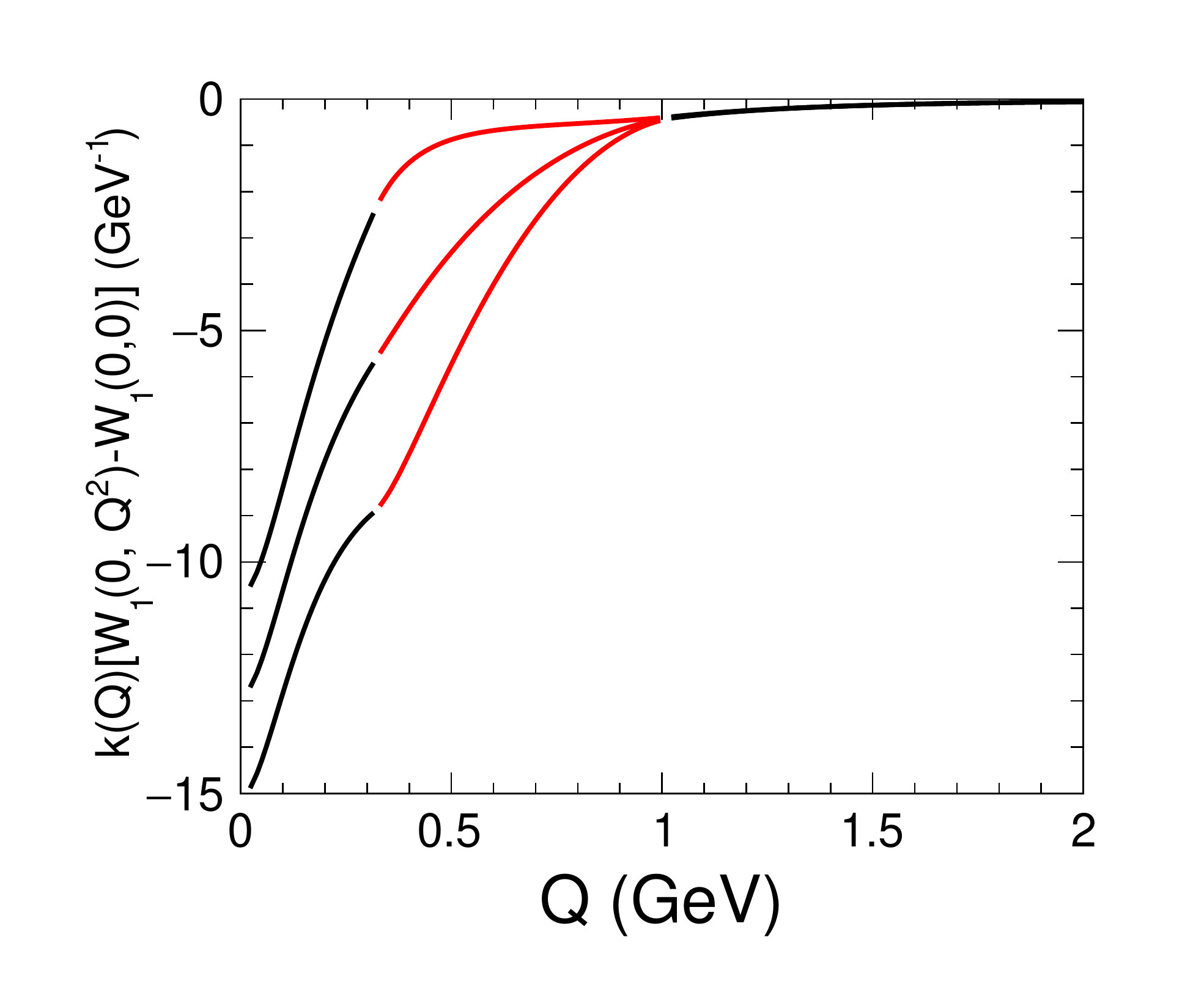}
  \vspace{-5mm}
  \caption{\label{fig:W1_TPE} Contribution of the subtraction function $W_1(0,Q^2)$
    to the
    two-photon exchange effect in the muonic hydrogen Lamb shift. 
    The black lines on the LHS and RHS of the plot show central value and
    error band for the low-$Q^2$ (NRQED) and high-$Q^2$ (OPE) regions. 
    The red lines in the intermediate region are interpolations. 
    The energy shift is proportional to the area under the curve.
    From Ref.~\cite{Hill:2016bjv}. 
  }
  \vspace{-5mm}
\end{figure} 

At high $Q^2 \gg m_p^2$, $W_1(0,Q^2)$ can be computed using operator
product expansion (OPE),
\begin{align}
 W_1(0,Q^2) &= {2m_p^2\over Q^2} A(Q^2) + \order(1/Q^4) \,, 
\end{align}
where $A(Q^2)$ is a product of perturbatively calculable coefficients
and nucleon matrix elements of local operators.  
The complete result for the leading OPE contains contributions from
both spin-0 and spin-2 operators.   Previous investigations of the impact
on the muonic hydrogen TPE have assumed subtraction functions without
the correct large-$Q^2$ behavior.
For example, Refs.~\cite{Birse:2012eb,Pachucki:1999zza}
consider an interpolation formula that identifies $A(Q^2)$ with low-energy
quantities. 
Reference~\cite{Miller:2012ne}
adopts a result in the literature~\cite{Collins:1978hi} for $A(Q^2)$
that contains the incorrect spin-0 contribution and entirely omits
the numerically dominant spin-2 contribution.
An extrapolation using for the first time the
correct low-$Q^2$ and high-$Q^2$ constraints
is displayed in Fig.~\ref{fig:W1_TPE}.

\begin{figure}[h]
  \centering
  \includegraphics[height=0.4\textwidth]{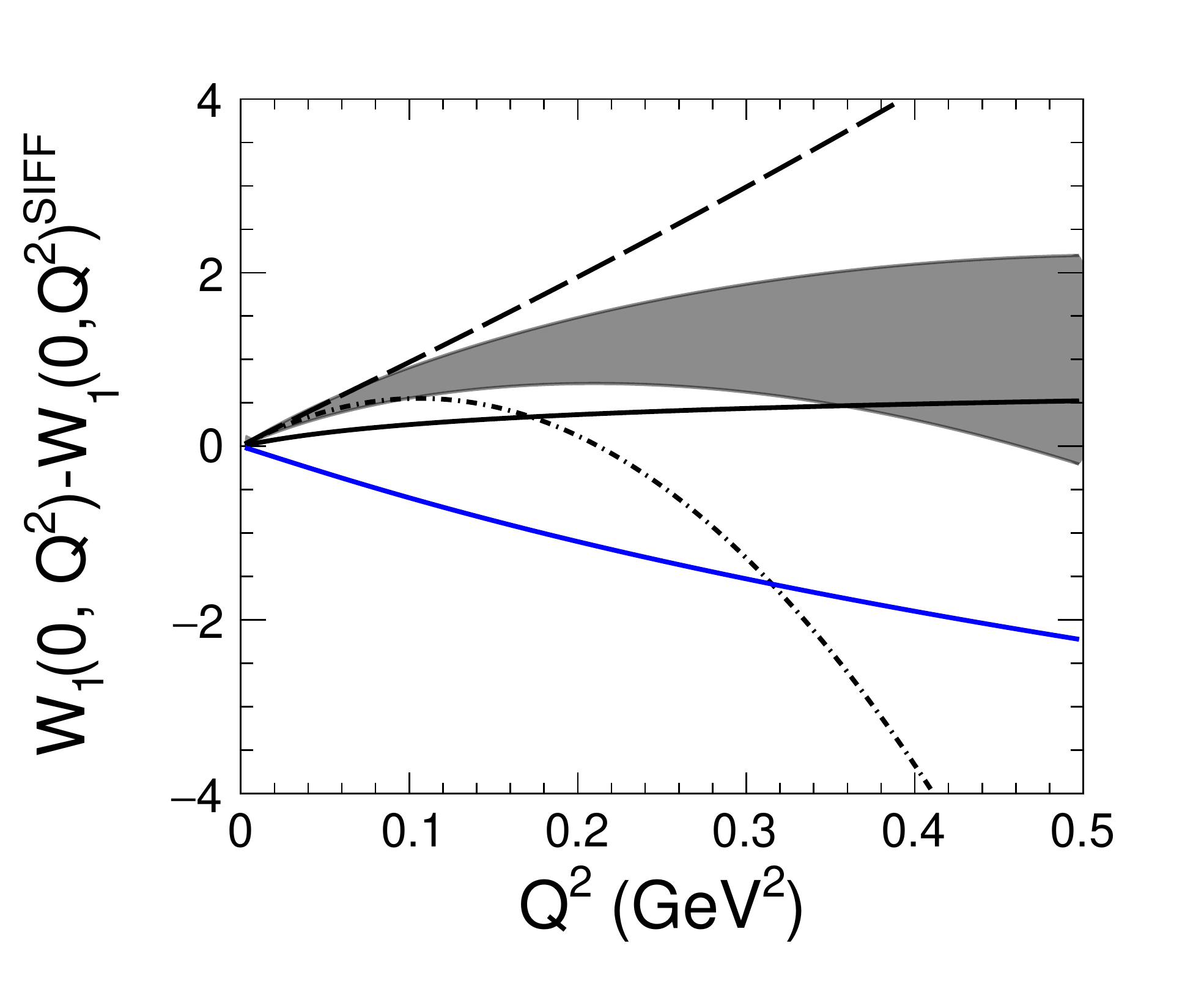}
  \includegraphics[height=0.4\textwidth]{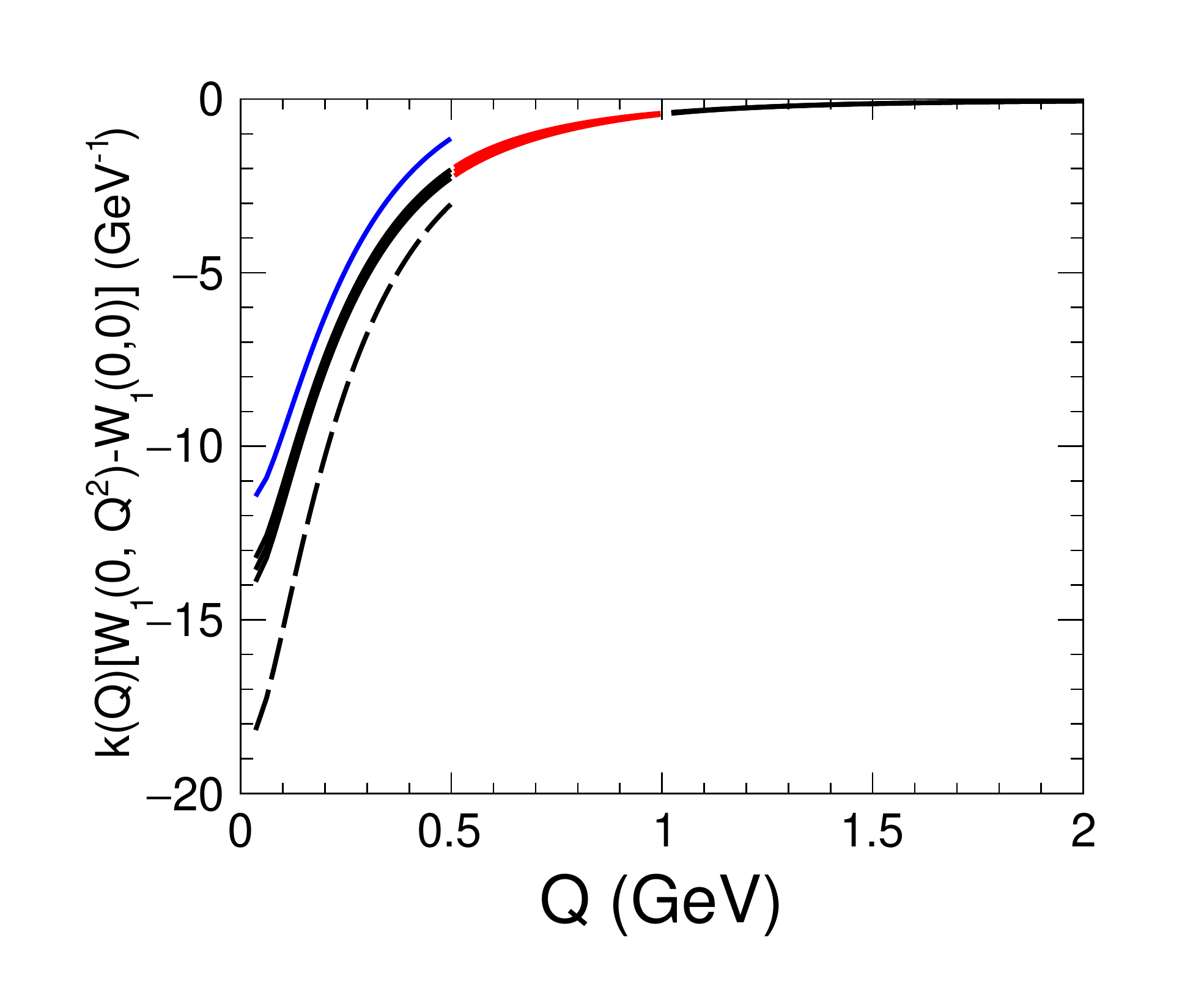}
  \vspace{-5mm}
  \caption{\label{fig:Birse} Contribution of subtraction function to two-photon exchange, using
    hadronic calculation of Birse and McGovern~\cite{Birse:2012eb}.  LHS: $W_1(0,Q^2)$
    using ``third order'', ``fourth order'' and ``fourth order plus $\Delta$'' results from
    Ref.~\cite{Birse:2012eb} (black solid line, black dashed line, solid band);
    ``fourth order plus $\Delta$'' Taylor expanded through $\order(Q^4)$ (dash-dotted black line);
    ``third order'' result from the relativistic formulation of Ref.~\cite{Alarcon:2013cba}.
    RHS: same as Fig.~\ref{fig:W1_TPE} but using the gray band on LHS for $W_1(0,Q^2)$
    [at fixed $W_1(0,Q^2)^{\rm SIFF}$].  Plots from Ref.~\cite{Hill:2016bjv}.
  }
  \vspace{-5mm}
\end{figure} 

Additional information on $W_1(0,Q^2)$ in the regime $m_\pi^2 \lesssim Q^2 \ll m_p^2$
can be obtained from chiral lagrangian analysis.
A sample result is displayed in Fig.\ref{fig:Birse}, where the LHS compares
different orders in the chiral expansion, and the RHS gives the analog
of Fig.~\ref{fig:W1_TPE}.
It should be noted that these curves perform a conventional separation of $W_1(0,Q^2)$
into a piece determined by elastic form factors (denoted by $W_1(0,Q^2)^{\rm SIFF}$
in Fig.~\ref{fig:Birse}, and for simplicity here assumed to have zero uncertainty),
and a remainder.
Uncertainties related to this separation are not contained in curves displayed
in Fig.~\ref{fig:Birse}. 

\begin{figure}[htb]
  \centering
  \psfragfig*[mode=nonstop,height=0.5\textwidth]{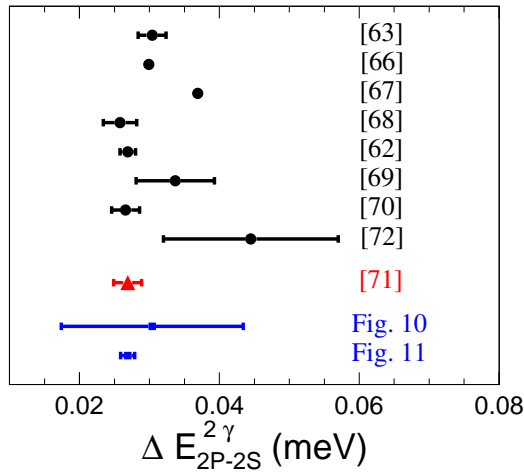}{
    \psfrag{a}{\cite{Pachucki:1999zza}}
    \psfrag{b}{\cite{Martynenko:2005rc}}
    \psfrag{c}{\cite{Nevado:2007dd}}
    \psfrag{d}{\cite{Carlson:2011zd}}
    \psfrag{e}{\cite{Birse:2012eb}}
    \psfrag{f}{\cite{Gorchtein:2013yga}}
    \psfrag{z}{\cite{Alarcon:2013cba}}
    \psfrag{h}{\color{blue}\hspace{-2mm} Fig.~\ref{fig:W1_TPE} }
    \psfrag{i}{\color{blue}\hspace{-2mm} Fig.~\ref{fig:Birse} }
    \psfrag{j}{\color{red}\cite{Antognini:2013jkc}}
    \psfrag{k}{\cite{Peset:2014jxa}}
  }
  \vspace{-5mm}
  \caption{Two-photon contribution to the Lamb shift in muonic hydrogen,
    adjusted to common proton elastic form factors (see text). 
    The red triangle is the summary of Ref.~\cite{Antognini:2013jkc} used
    in the 2013 CREMA muonic hydrogen extraction of $r_E^p$.  Plot from Ref.~\cite{Hill:2016bjv}.
    \label{fig:collect}}
\end{figure}

Fig.~\ref{fig:collect} displays results in the literature for
the TPE contribution to the muonic hydrogen Lamb shift, including
the results corresponding to Figs.~\ref{fig:W1_TPE} and \ref{fig:Birse}. 
The TPE contribution remains a dominant source of uncertainty on the
$r_E^p$ extraction from muonic hydrogen. 

\section{New experimental clues \label{sec:expt}}

There are several directions of activity to better understand
the proton radius puzzle.

\subsection{New muonic atom measurements}

The Lamb shift in muonic deuterium has been measured in Ref.~\cite{Pohl1:2016xoo}.
Combined with the regular hydrogen-deuterium isotope shift of the 1S-2S
transition~\cite{Parthey:2010aya}, this result
has been translated to a value
\begin{align}
  r_E^p(\mu D) &= 0.83562(21) \,{\rm fm} \,. 
\end{align}
Further new measurements are anticipated~\cite{Antognini:2011zz} in muonic ${}^3{\rm He}$ and ${}^4{\rm He}$,
where nuclear structure effects are important for
interpretation~\cite{Dinur:2015vzv,Hernandez:2016jlh,Carlson:2016cii}.

\subsection{New regular hydrogen measurements}

It may be noted that no single measurement in regular hydrogen differs from 
the muonic hydrogen line in Fig.~\ref{fig:2010} by more than $~\sim 2\sigma$, 
raising the question of how to properly average these results.%
\footnote{
  For a discussion, see Ref.~\cite{Beyer:2013jla}. 
}
New preliminary results for the hydrogen $2S-4P$ splitting 
have been reported by Beyer et al.~\cite{Beyer:2013jla,Beyer}, with a
``small'' radius, and error comparable to the existing hydrogen average.
In the absence of a final result, Fig.~\ref{fig:prospects} displays the
anticipated radius sensitivity,%
\footnote{
  An uncertainty $0.010\,{\rm fm}$ has been assumed in the figure. 
}
with central value taken for illustration
as the CODATA 2014 electron combination. 
For an overview of other potential regular hydrogen measurements
see Ref.~\cite{2016EPJWC.11301006A}.  A range of new measurements with
hydrogen-like He$^+$ ions~\cite{2009PhRvA..79e2505H,2010PhRvL.105f3001K,2014NatPh..10...30M}, molecules and molecular
ions~\cite{2009JChPh.130q4306L,2014PhRvL.113b3004S,2013PhRvL.110s3601D,2016NatCo...710385B,Karr:2016bwq}, and circular
Rydberg states~\cite{Jentschura:2008zz,2011PhST..144a4009T} are also anticipated.%
\footnote{For further discussion see Ref.~\cite{2016arXiv160703165P}.}

\subsection{Low-$Q^2$ electron-proton scattering}

Form factor nonlinearities can be theoretically controlled, provided
that experimental errors and correlations are precisely specified.
However, it is interesting to extract the proton charge radius
entirely from low-$Q^2$ data, especially given the apparent tension
between low-$Q^2$ and high-$Q^2$ data illustrated in
Fig.~\ref{fig:Q2max}.  The PRad experiment at
JLab~\cite{Gasparian:2014rna} utilizes a non-magnetic spectrometer, a
windowless target, and a simultaneous calibration with M\o ller
scattering to control experimental systematics.  First data was
collected in May/June 2016, and a first analysis is anticipated in
2017.  Fig.~\ref{fig:prospects} displays the anticipated radius
sensitivity,%
\footnote{ A 1\% radius uncertainty is assumed in the figure.  }
with central value taken for definiteness as the CODATA 2014 electron combination. 
Low-$Q^2$ data have also been collected using the initial state
radiation (ISR) technique at MAMI~\cite{Mihovilovic:2016rkr}.  An
experiment with a new target is being planned that will reduce backgrounds
and access $Q^2 \sim 10^{-4}\,{\rm GeV}^2$. 

\subsection{Muon-proton scattering}

\begin{figure}[h]
  \centering
    \hspace{-10mm}
    \psfragfig*[mode=nonstop,height=1.1\textwidth]{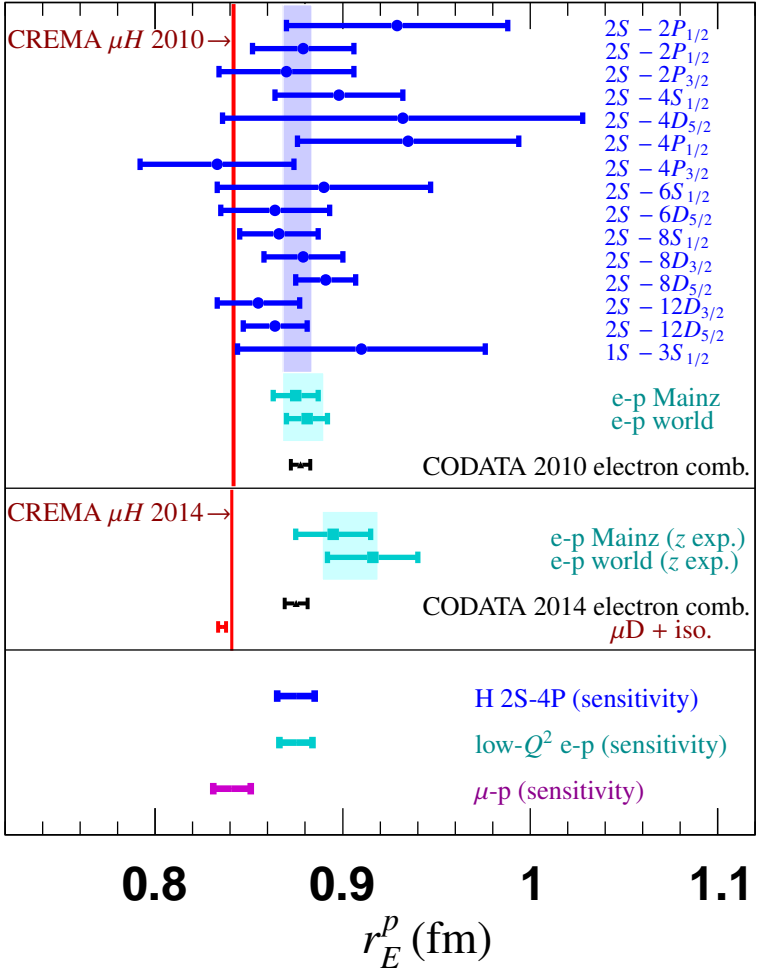}{
    \psfrag{a}{\small\color{blue}{$2S-2P_{1/2}$} }
    \psfrag{b}{\small\color{blue}{$2S-2P_{1/2}$} }
    \psfrag{c}{\small\color{blue}{$2S-2P_{3/2}$} }
    \psfrag{d}{\small\color{blue}{$2S-4S_{1/2}$} }
    \psfrag{e}{\small\color{blue}{$2S-4D_{5/2}$} }
    \psfrag{f}{\small\color{blue}{$2S-4P_{1/2}$} }
    \psfrag{g}{\small\color{blue}{$2S-4P_{3/2}$} }
    \psfrag{h}{\small\color{blue}{$2S-6S_{1/2}$} }
    \psfrag{i}{\small\color{blue}{$2S-6D_{5/2}$} }
    \psfrag{j}{\small\color{blue}{$2S-8S_{1/2}$} }
    \psfrag{k}{\small\color{blue}{$2S-8D_{3/2}$} }
    \psfrag{l}{\small\color{blue}{$2S-8D_{5/2}$} }
    \psfrag{m}{\small\color{blue}{$2S-12D_{3/2}$} }
    \psfrag{n}{\small\color{blue}{$2S-12D_{5/2}$} }
    \psfrag{o}{\small\color{blue}{$1S-3S_{1/2}$} }
    \psfrag{p}{\color{darkcyan}{\hspace{0mm} e-p Mainz
    }}
    \psfrag{q}{\color{darkcyan}{\hspace{0mm} e-p world
    }}    
    \psfrag{r}{\color{black}{\hspace{-25mm} CODATA 2010 electron comb.
    }}
    \psfrag{s}{\color{darkcyan}{\hspace{-8mm} e-p Mainz ($z$ exp.)
    }}
    \psfrag{t}{\color{darkcyan}{\hspace{-8mm} e-p world ($z$ exp.)
    }}    
    \psfrag{u}{\color{black}{\hspace{-25mm} CODATA 2014 electron comb.
    }}
    \psfrag{v}{\color{darkred}{\hspace{0mm} $\mu$D + iso.
    }}
    \psfrag{w}{\color{blue}{\hspace{-18mm} H 2S-4P (sensitivity)
    }}
    \psfrag{z}{\color{darkcyan}{\hspace{-18mm} low-$Q^2$ e-p (sensitivity)
    }}
    \psfrag{y}{\color{darkmagenta}{\hspace{-18mm} $\mu$-p (sensitivity)
    }}    
    \psfrag{x}{\LARGE \color{black}{$r_E^p$\,(\rm fm)}}
    \psfrag{xx}{ \color{darkred}{\hspace{-10.5mm}CREMA $\mu H$ 2010\! $\rightarrow$} }
    \psfrag{yy}{ \color{darkred}{\hspace{-10.5mm}CREMA $\mu H$ 2014\! $\rightarrow$} }
  }
  \vspace{-15mm}
  \caption{
    \label{fig:prospects}
    Status of the proton radius puzzle circa 2016, with prospects for new data.
    The upper pane is reproduced from Fig.~\ref{fig:2010}.
    The middle pane shows updated results.  The cyan points give updated fits to electron scattering
    data using $z$ expansion (final two points in Fig.~\ref{fig:reanalysis}, from Ref.~\cite{Lee:2015jqa}.
    The black point represents the 2014 CODATA~\cite{Mohr:2015ccw} combination of hydrogen and
    electron-proton scattering determinations.
    The red point is from the 2016 CREMA muonic deuterium Lamb shift measurement using the
    regular hydrogen-deuterium isotope shift~\cite{Pohl1:2016xoo}.
    The bottom pane shows expected sensitivities of anticipated results in: regular hydrogen~\cite{Beyer} (blue);
    low-$Q^2$ electron-proton scattering~\cite{Gasparian:2014rna} (cyan);
    and muon-proton scattering~\cite{Gilman:2013eiv} (magenta).  See text for details. 
  }
  \vspace{-5mm}
\end{figure}

As illustrated in Fig.~\ref{fig:2010}, both bound state and scattering
measurements exist for the electron system.   Only a bound state
measurement exists for the muon system.  A measurement of the proton
charge radius  from muon-proton elastic scattering has been proposed
by the MUSE collaboration~\cite{Gilman:2013eiv}.  A combination of
$e^\pm$ and $\mu^\pm$ data should provide cancellation of systematic
errors and empirical constraints on two-photon exchange.
Fig.~\ref{fig:prospects} displays the anticipated radius sensitivity,%
\footnote{
  An uncertainty $0.010\,{\rm fm}$ for the radius determined from muon
  scattering has been assumed in the figure.  The difference between
  radii from electron and muon scattering should have smaller uncertainty.
}
with central value taken for illustration as the previous muonic
hydrogen average. 

\subsection{Summary of status and prospects}

Figure~\ref{fig:prospects} displays the current status of the proton
radius puzzle.  Compared to Fig.~\ref{fig:2010}, the muonic hydrogen error bar
has been increased to reflect updates and a revised treatment of
TPE in Ref.~\cite{Antognini:2013jkc}, and
the new muonic deuterium data point has been included. 
The electron scattering results reflect the treatment of form factor
nonlinearities and more conservative systematic errors from
Ref.~\cite{Lee:2015jqa}.
In addition to these existing results, projected sensitivities
are illustrated for the $2S-4P$ measurement in regular hydrogen; low-$Q^2$ electron-proton
scattering; and muon-proton scattering.

\section{Some broader implications \label{sec:broader}}

Regardless of its resolution, the proton radius
puzzle has acted as a ``disruptive technology'',
highlighting critical areas where better understanding
is needed, ranging from precision measurements
and fundamental constants to neutrino physics and formal questions in
effective field theory.

\subsection{Nucleon form factors for neutrinos and other processes}

\begin{figure}[htb]
  \centering
  \includegraphics[height=0.5\textwidth]{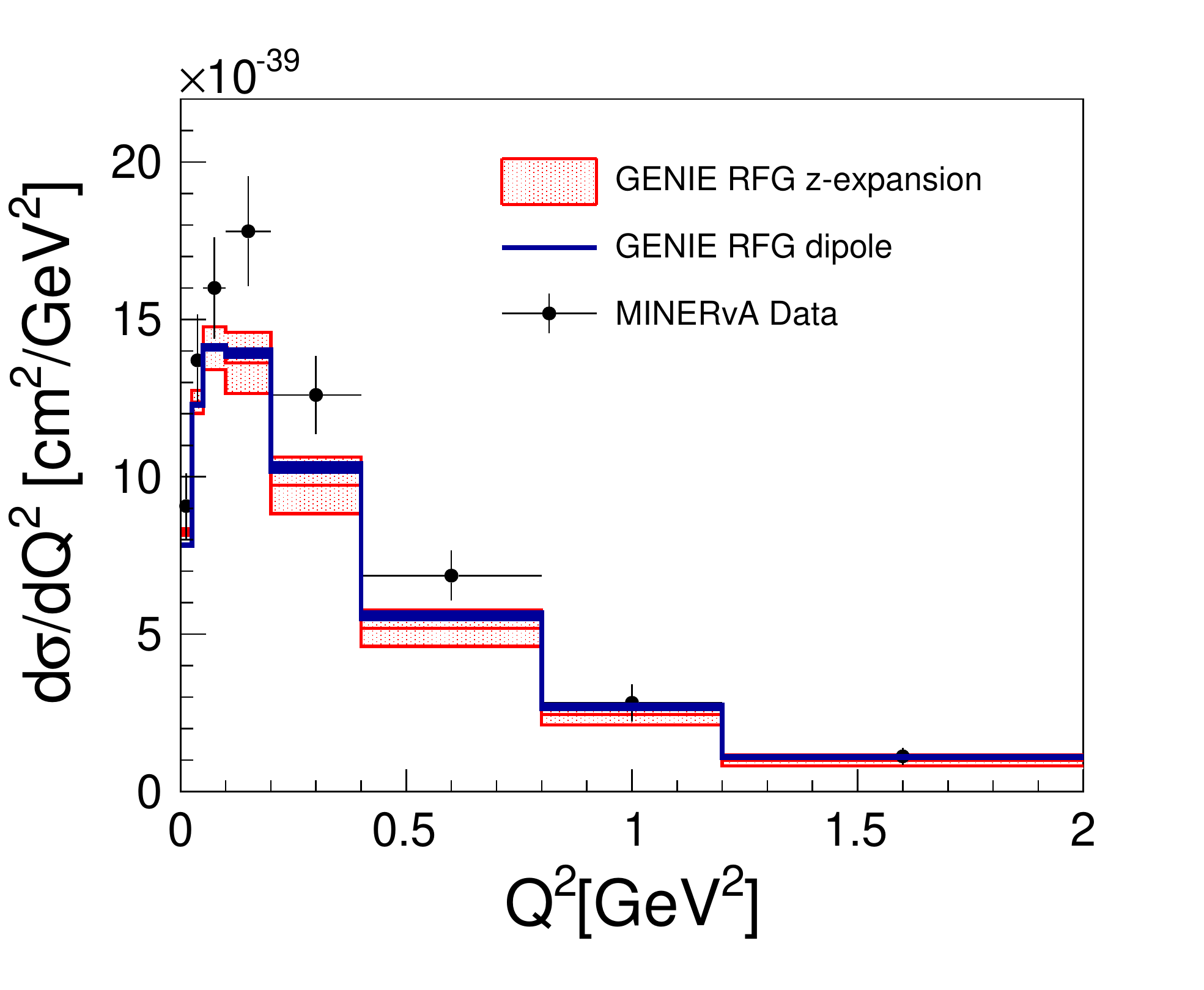}
  \vspace{-5mm}
  \caption{\label{fig:neutrino} Comparison of experimental data from
    the MINERvA collaboration~\cite{Fiorentini:2013ezn} with the
    theory prediction in a simple nuclear model, showing error induced by
    nucleon-level form factors.  The previously underestimated uncertainty
    is in blue, the updated uncertainty is in red.  From Ref.~\cite{Meyer:2016oeg}. 
  }
\end{figure} 

The elastic form factors of the nucleon are precisely defined quantities
that impact many observables.  The proton radius puzzle has emphasized
the importance of properly accounting for form factor shape uncertainty
and radiative corrections in the analysis of experimental data.  
An important application of current relevance is neutrino-nucleus
scattering at long baseline oscillation experiments.%
\footnote{
  Another recent application is to the study of photon initiated processes
  at the LHC~\cite{Manohar:2016nzj}.  
  }
At the nucleon-level, the basic signal process for neutrino detection
is charged-current quasielastic scattering, $\nu_\ell n \to \ell^- p$.
Figure~\ref{fig:neutrino} illustrates that with proper treatment of
form factor shape, the nucleon-level uncertainty on the cross section is an order
of magnitude larger than previously thought.%
\footnote{
  Reference~\cite{Meyer:2016oeg} determines $r_A^2=0.46(22)\,{\rm fm}^2$ from existing data.  
  Other results in the literature quote much smaller uncertainty, for example
  Ref.~\cite{Bodek:2007ym} finds $r_A^2=0.454(12)\,{\rm fm}^2$.
  This small uncertainty arises from a dipole shape assumption that
  was applied to neutrino scattering and pion electroproduction data~\cite{Bodek:2007ym}.
  Taken at face value, the uncertainty
  on the axial radius would be {\it smaller} than the uncertainty on
  scattering determinations of the proton
  electric charge radius, even though the former results from small statistics
  neutrino beams with poorly known flux on nuclear targets (vs. high statistics
  electron beams of monoenergetic energy on proton targets).  
  }
This complicates the program of constraining nuclear effects from
measurement: for example, in Fig.\ref{fig:neutrino}, the data-theory
discrepancy is due to a combination of nucleon-level input uncertainty
and nuclear modeling error.
As a quantitative benchmark, the $\sim 10\%$ uncertainty on the nucleon-level
cross section, $\sigma(\nu_\mu n \to \mu^- p)|_{E_\nu=1\,{\rm GeV}} = 10.1(0.9)\times 10^{-39}\,{\rm cm}^2$
due to axial form factor shape uncertainty already saturates the error budget
of next generation experiments.%
\footnote{
  Determination of the requisite nuclear corrections presently
  relies on data-driven
  modeling~\cite{Andreopoulos:2009rq,Hayato:2009zz,Buss:2011mx,Golan:2012wx,Katori:2016yel}
  employing experimental
  constraints~\cite{Drakoulakos:2004gn,AguilarArevalo:2010zc,AguilarArevalo:2010bm,AguilarArevalo:2010cx,Abe:2015awa,Anderson:2011ce,Rodrigues:2015hik}.
  Ab-initio nuclear computations are beginning to provide additional
  insight~\cite{Lovato:2013cua,Bacca:2014tla,Carlson:2014vla}.
  Regardless of whether nuclear corrections are constrained
  experimentally or derived from first principles, independent knowledge
  of the elementary nucleon-level amplitudes is essential.
}
This situation motivates an improvement of elementary nucleon-level amplitudes
from lattice QCD~\cite{Meyer:2016kwb,Bhattacharya:2016zcn,Yoon:2016jzj},
from new precise neutrino data, 
and/or potential advances in other fields such as muon capture in muonic
hydrogen~\cite{Andreev:2012fj}. 
More refined treatment of QED radiative corrections is also needed for
neutrino experiments~\cite{Day:2012gb,Hill:2016gdf}. 

\subsection{Formal questions in effective field theory}

Certain radiative corrections to nuclear structure effects contributing
to muonic hydrogen demand consideration of the $1/M^4$ heavy particle
lagrangian (NRQED)~\cite{Hill:2012rh}.
It is only at this order that an interplay of Lorentz and gauge
symmetry causes a violation~\cite{Heinonen:2012km}
of reparameterization invariance as
implemented by a classic ansatz~\cite{Luke:1992cs}. 
Muonic hydrogen helped uncover this interesting
feature of effective field theory. 

\subsection{Nonperturbative methods}

Lattice QCD offers a route to nucleon form factors that is
independent of detector-dependent radiative corrections
(although the impact of QED corrections to lattice predictions
must be robustly estimated).
In the context of the axial form factor, probed most directly
in neutrino scattering, practical considerations (underground
safety for hydrogen or deuterium targets) may also impede
further experimental progress.
Next generation lattice QCD is poised to
contribute.%
\footnote{
  For recent work, see: 
  \cite{Bhattacharya:2016zcn,Meyer:2016kwb,Yoon:2016jzj,Abramczyk:2016ziv,Yamazaki:2015vjn,Liang:2016fgy,Abdel-Rehim:2015owa,Djukanovic:2016ocj}.
}
There are also novel methods under investigation to access the radii directly
on the lattice, instead of using the form factor as
intermediary~\cite{Alexandrou:2016hiy}. 

The precision spectroscopy of muonic atoms heavier than hydrogen
demands correspondingly precise 
nuclear structure calculations on light nuclei, and is
a proving grounds for ab initio nuclear
methods~\cite{Machleidt:2000ge,Friar:2013rha,Hernandez:2014pwa,Pachucki:2015uga}. 
There is presently reasonable agreement between different
methods for deuteron structure corrections, although (cf. Fig.~\ref{fig:prospects})
there remains an intriguing $\sim 2.5\sigma$ discrepancy between the
deuteron radius measured in regular and muonic deuterium. 

\subsection{Fundamental constants} 

Taking the muonic hydrogen results
at face value implies a shift in the Rydberg constant by
$\sim 7\sigma$~\cite{Mohr:2012tt,Antognini:1900ns}. 
The proton charge radius is a well defined observable of Nature,%
\footnote{
  This observable is of course determined by the parameters appearing
  in the Standard Model lagrangian, by a relation that we cannot
  yet determine with the accuracy of the muonic hydrogen measurement. 
}
and again taking the muonic hydrogen result at face value, this
observable will change by $\sim 7\sigma$ (correlated with the Rydberg).

\subsection{Motivating searches for new phenomena}

The proton radius puzzle has motivated a variety of
investigations into possible phenomena beyond the Standard
Model~\cite{Barger:2010aj,TuckerSmith:2010ra,Batell:2011qq,Carlson:2012pc,Carlson:2013mya,Pauk:2015oaa,Liu:2016qwd},
part of the broader program of searching for violations of lepton universality
and light, weakly-coupled new physics scenarios~\cite{Bryman:2011zz,redtop}. 
The puzzle has also motivated novel perspectives on computations within
the Standard Model; for a recent example see
Refs.~\cite{Burgess:2016zbp,Burgess:2016lal,Burgess:2016ddi}.

\section{Outlook \label{sec:outlook}} 

Laser spectroscopy of light muonic atoms has acted as a disruptive
technology, demanding and motivating the
development of better theoretical tools, and
instigating a range of new experimental measurements.

The extraction of the proton charge radius from electron scattering
data remains a controversial topic.
Within the quoted experimental uncertainties of the highest
statistics dataset~\cite{Bernauer:2013tpr}, and employing
standard models for radiative corrections, the data prefer
a value of $r_E^p$ significantly higher than muonic
hydrogen (cf. Fig.~\ref{fig:reanalysis}).  
It is straightforward to account for form factor nonlinearities
(the proper treatment of these effects is essential, cf. the
alarming discrepancy between the first two points in Fig.~\ref{fig:reanalysis}). 
At the same time it is critical to account for radiative
corrections, both soft (model independent) and hard
(model dependent).
A complete calculation of the Sudakov-enhanced soft corrections
(cf. Fig.~\ref{fig:radcor})
reveals deficiencies in previous treatments that are in tension
with the assumed experimental error budget~\cite{Bernauer:2013tpr},
but appear unlikely to solely account for the radius anomaly.
The same calculation reveals an ambiguity in the treatment of
hard TPE of similar magnitude. 
Of special interest are new measurements that can
address remaining theoretical uncertainties,
like hard two-photon exchange~\cite{Henderson:2016dea}.

New electron scattering measurements focused on low $Q^2$
are anticipated to provide an alternative determination of
$r_E^p$.  While of fundamental importance, such measurements
will not by themselves shed light on whether underestimated systematic effects
are impacting the higher-$Q^2$ data (and whether such effects
can be ignored in the low-$Q^2$ data).  
Similarly, several theoretical approaches employ a subset of
the scattering data (typically at low-$Q^2$),
and/or effectively overrule the scattering data with other
constraints to find a charge radius consistent with muonic
hydrogen.  
Irrespective of details, such approaches cannot
by themselves explain the existing anomaly
(which requires consideration of a
broad $Q^2$ range, cf. Fig.~\ref{fig:sens}), 
and cannot offer a satisfactory resolution to the proton
radius puzzle.  

This point becomes especially acute for the analog process
of neutrino scattering,
where both form factor shape and radiative corrections
play an important role, and where quantitative control of
cross sections over a broad range of $Q^2$ is critical
for the global long baseline neutrino program.  Here we
do not have the luxury of restricting to small $Q^2$,
or to enforcing constraints on spectral functions using
the analog of precise $e^+ e^-$ hadron production data.  

The structure-dependent TPE effect in muonic hydrogen
has received significant attention from a variety
of approaches (cf. Fig.~\ref{fig:collect}).
This contribution remains a dominant uncertainty in the
determination of the $r_E^p$ from muonic hydrogen, but
new constraints on the subtraction function leave limited
room for large effects
(cf. Fig.~\ref{fig:W1_TPE}, and Fig.~\ref{fig:Birse} RHS). 

The proton radius puzzle
has motivated investigations leading to new
formal results in effective field theory, 
surprises in seemingly well-established
operator product expansions, and many new
analyses of hadronic structure using a variety
of effective field theories.  
New experimental efforts in the realm of lepton
universality tests are coming online and promise
yet further insights.  

\vskip 0.15in
\noindent
{\bf Acknowledgments:}
I am grateful to the organizers of Confinement 2016 for a stimulating
meeting, and to many collaborators, J.~Arrington, M.~Betancournt, 
B.~Bhattacharya, R.~Gran, J.~Heinonen, A.~Kronfeld, R.~Li, G.~Lee, A.~Meyer, 
G.~Paz, J.~Simone, M.~Solon, Z.~Ye on topics connected with the proton radius
puzzle, as well as many other colleagues for insightful discussion,
especially A.~Gasparian, R.~Gilman, M.~Mihovilovic and R.~Pohl for communicating the
status of prospective measurements. 
This work was supported by NIST Precision Measurement Grants Program
and DOE grant DE-FG02-13ER41958.  Research at Perimeter Institute is
supported by the Government of Canada through the Department of
Innovation, Science and Economic Development and by the Province of
Ontario through the Ministry of Research and Innovation.  TRIUMF
receives federal funding via a contribution agreement with the
National Research Council of Canada.  Fermilab is operated by Fermi
Research Alliance, LLC under Contract No. DE-AC02-07CH11359 with the
United States Department of Energy.

\end{document}